\documentclass[
 reprint,
 superscriptaddress,
 amsmath,amssymb,
 aps,
 pra,
]{revtex4-2}

\usepackage{graphicx}   
\usepackage{caption}
\usepackage{subcaption}
\usepackage{dcolumn}    
\usepackage{bm}         
\usepackage{xcolor}
\usepackage[colorlinks=true,
            allcolors=red
           ]{hyperref}
\usepackage{braket}
\usepackage{comment}
\newcommand{\be}{\begin{equation}}
\newcommand{\ee}{\end{equation}}
\newcommand{\bea}{\begin{align}}
\newcommand{\eea}{\end{align}}

\begin{document}
\title{Extending QAOA-GPT to Higher-Order Quantum Optimization Problems}

\author{Leanto Sunny}
\email{lsunny@vols.utk.edu}
\affiliation{Department of Physics and Astronomy, The University of Tennessee, Knoxville, TN 37996-1200, USA}

\author{Abhinav Rijal}
\email{arijal@vols.utk.edu}
\affiliation{Department of Physics and Astronomy, The University of Tennessee, Knoxville, TN 37996-1200, USA}

\author{George Siopsis}
\email{siopsis@tennessee.edu}
\affiliation{Department of Physics and Astronomy, The University of Tennessee, Knoxville, TN 37996-1200, USA}

\date{\today}

\begin{abstract}
The recently proposed QAOA-GPT framework \cite{tyagin2025qaoa} demonstrated that generative pre-trained transformers can learn mappings between problem graphs and optimized quantum circuits for the Quantum Approximate Optimization Algorithm (QAOA). In this work, we extend QAOA-GPT to Higher-Order Unconstrained Binary Optimization (HUBO) problems, focusing on spin-glass Hamiltonians that include cubic interaction terms. Using FEATHER graph embeddings to encode topological information, we train the model on graph–circuit pairs generated via ADAPT-QAOA and evaluate its performance on 8- and 16-qubit instances embedded on heavy-hex lattices. The generative model produces adaptive QAOA-like circuits and corresponding variational parameters in a single forward pass, bypassing the iterative classical optimization loop. The generated circuits achieve average approximation ratios exceeding 0.95, closely matching classically optimized ADAPT-QAOA results, while maintaining consistent parameter distributions across circuit depths. These results demonstrate that QAOA-GPT generalizes effectively to higher-order cost Hamiltonians and complex energy landscapes, establishing generative modeling as a scalable pathway toward autonomous variational circuit design and quantum algorithm discovery in the NISQ era.
\end{abstract}

\maketitle

\section{Introduction} \label{sec:intro}

Quantum computing promises to transform the solution of complex problems in optimization, chemistry, and materials science by exploiting the superposition and entanglement of quantum states to explore exponentially large configuration spaces. In the near term, variational quantum algorithms (VQAs) have emerged as a practical strategy for leveraging noisy intermediate-scale quantum (NISQ) devices \cite{cerezo2021variational}. By combining a parameterized quantum circuit \cite{PhysRevResearch.3.023092} with a classical optimizer in a hybrid feedback loop, VQAs prepare quantum states that approximately minimize an objective function derived from a problem Hamiltonian. Prominent examples include the Variational Quantum Eigensolver (VQE) for molecular energy estimation and the Quantum Approximate Optimization Algorithm (QAOA) for combinatorial optimization problems \cite{farhi2014quantumapproximateoptimizationalgorithm}.

QAOA constructs a variational ansatz through alternating applications of a cost Hamiltonian, encoding the problem instance, and a mixer Hamiltonian, which drives transitions between computational basis states. The variational parameters governing these layers are optimized to minimize the expectation value of the cost Hamiltonian. Despite its conceptual simplicity and hardware compatibility, QAOA remains limited by the difficulty of identifying optimal parameters, the exponential growth of the search space with circuit depth, and the hardware noise sensitivity of long circuits \cite{blekos2024review}. Adaptive approaches such as ADAPT-QAOA \cite{zhu2022adaptqaoa} have sought to address these limitations by building circuits iteratively, adding operators that yield the steepest local gradient descent in energy. While this adaptive strategy produces more expressive and compact ansätze, it incurs significant computational cost due to repeated gradient evaluations and classical optimization.

This has motivated data-driven strategies to learn these complex optimization patterns. Initial approaches have used Graph Neural Networks to predict QAOA parameters \cite{jain2022graph}, but these often lack the generative capacity for full circuit synthesis. Recently, Tyagin \textit{et al.}\ introduced the QAOA-GPT framework \cite{tyagin2025qaoa}, which reimagines QAOA circuit construction as a sequence-generation problem. Using a generative pre-trained transformer (GPT) trained on graph–circuit pairs, QAOA-GPT learns correlations between problem structure and optimized variational parameters. The trained model can then generate efficient QAOA circuits for unseen graph instances in a single forward pass, eliminating the need for iterative classical optimization and dramatically reducing computation time. The framework demonstrated strong performance for quadratic unconstrained binary optimization (QUBO) problems such as MaxCut, producing circuits with accuracy comparable to ADAPT-QAOA while maintaining consistent scalability.

In this work, we extend the QAOA-GPT paradigm to Higher-Order Unconstrained Binary Optimization (HUBO) problems, focusing on spin-glass Hamiltonians that include cubic interaction terms. These higher-order systems possess rugged and frustrated energy landscapes, making them an ideal benchmark for testing the generalization capacity of generative quantum-circuit models. We employ FEATHER graph embeddings \cite{rozemberczki2020characteristic} to encode topological and higher-order connectivity information into the model input and generate datasets using ADAPT-QAOA-optimized circuits for 8- and 16-qubit spin-glass instances embedded on heavy-hex graphs. The resulting model, trained end-to-end on these graph–circuit pairs, autonomously produces adaptive QAOA-like circuits and corresponding variational parameters without any classical post-optimization.

We evaluate the performance of the extended QAOA-GPT framework on unseen HUBO instances and analyze its learned parameter distributions. The model achieves average approximation ratios exceeding 0.95 for 16-qubit systems, matching ADAPT-QAOA benchmarks while offering orders-of-magnitude faster inference. The smooth scaling of parameters with circuit depth further confirms that the generative model internalizes the structure of effective QAOA ans\"atze.

This work demonstrates that generative modeling can serve as a scalable and data-driven mechanism for quantum circuit synthesis, extending beyond quadratic optimization to higher-order and more complex Hamiltonians. Beyond its immediate application to spin-glass models, the present study establishes a foundation for autonomous discovery of problem-specific quantum algorithms—an essential step toward realizing practical quantum advantage in the NISQ regime.


This paper is organized as follows. Section \ref{sec:method} provides a detailed overview of the theoretical foundations of QAOA and graph embedding via FEATHER \cite{rozemberczki2020characteristic}, along with the inner workings of the QAOA-GPT model and the experimental setup, including the simulation implementations used in this study. Section \ref{sec:results} presents our experimental results, comparing the performance of our GPT-based approach against classical methods and ADAPT-QAOA. Finally, Section \ref{sec:conclusions} discusses the implications of our findings and outlines promising future research directions for QAOA-GPT frameworks.

\section{Background \& Methodology} \label{sec:method}

This section outlines the theoretical and computational foundations underlying our study.
We begin by briefly reviewing QAOA and its adaptive variant (ADAPT-QAOA), which provide the variational framework upon which our generative model is built.
We then describe the FEATHER graph-embedding method used to encode structural information about problem instances, followed by the architecture and training procedure of the QAOA-GPT model employed in this work.
Finally, we define the higher-order spin-glass Hamiltonians used as benchmarks for evaluating the model’s performance.
Together, these subsections establish the methodological basis for extending QAOA-GPT to  HUBO problems.

\subsection{QAOA}

QAOA \cite{farhi2014quantumapproximateoptimizationalgorithm} is a hybrid quantum–classical framework designed to find approximate solutions to discrete combinatorial optimization problems, especially those that can be mapped to Ising Hamiltonians. Given a problem represented on a graph 
$G(V,E)$, the cost Hamiltonian 
$H_C$ encodes the objective function whose ground state corresponds to the optimal solution, while a mixer Hamiltonian 
$H_M$ drives transitions between computational basis states to explore the solution space. For a graph with weighted edges 
$w_ij$, these operators typically take the form
    \begin{equation}
        H_C = \sum_{(i,j) \in E} w_{ij} Z_i Z_j, \ H_M = \sum_i X_i,
    \label{mixer}
    \end{equation}
    where $Z_i$ and $X_i$ denote Pauli operators acting on qubit $i$.

Starting from the uniform superposition  $\ket{+}^{\otimes n}$, QAOA alternates between cost and mixer unitaries for $p$ layers:
\begin{equation}
    \ket{\psi(\boldsymbol{\gamma}, \boldsymbol{\beta})}
    = \prod_{k=1}^{p} e^{-i \beta_k H_M} e^{-i \gamma_k H_C} \ket{+}^{\otimes n},
    \label{eq:qaoa_state}
\end{equation}
where $\boldsymbol{\gamma} = (\gamma_1, \ldots, \gamma_p)$ and $\boldsymbol{\beta} = (\beta_1, \ldots, \beta_p)$ are sets of variational parameters.

The objective function is defined as
\begin{equation}
    C(\boldsymbol{\gamma}, \boldsymbol{\beta}) = 
    \bra{\psi(\boldsymbol{\gamma}, \boldsymbol{\beta})} H_C \ket{\psi(\boldsymbol{\gamma}, \boldsymbol{\beta})},
\end{equation}
which is minimized by a classical optimizer to approximate the ground-state energy of $H_C$.

QAOA is compatible with near-term quantum hardware due to its shallow circuit structure, but its efficiency critically depends on the choice of parameters 
$\boldsymbol{\gamma}, \boldsymbol{\beta}$. Exhaustive or gradient-based searches over these parameters scale poorly with system size, motivating data-driven or adaptive alternatives.

\subsection{ADAPT-QAOA}

The ADAPT-QAOA approach \cite{zhu2022adaptqaoa} enhances expressivity by adaptively constructing the variational circuit. Instead of fixing the ansatz structure, the algorithm selects, at each iteration, the operator from a predefined pool $\mathcal{A} = \{A_1, A_2, \ldots, A_N\}$ that produces the steepest energy descent. The gradient of the cost function with respect to each candidate operator $A_j$ is computed as
\begin{equation}
    g_j 
    = -i \, \bra{\psi^{(k-1)}} 
        e^{i\gamma_0 H_C} [H_C, A_j] e^{-i\gamma_0 H_C}
      \ket{\psi^{(k-1)}},
    \label{eq:adaptqaoa_gradient}
\end{equation}
where $\ket{\psi^{(k-1)}}$ is the current variational state.

The operator $A^{(k)}$ corresponding to the largest $|g_j|$ is appended to the circuit, yielding the update
\begin{equation}
    \ket{\psi^{(k)}} 
    = e^{-i \beta_k A^{(k)}} e^{-i \gamma_k H_C} \ket{\psi^{(k-1)}}.
    \label{eq:adaptqaoa_layer}
\end{equation}
This iterative procedure continues until convergence or a predefined circuit depth is reached.

Although ADAPT-QAOA generates compact, problem-tailored circuits with high approximation quality, it remains computationally demanding because each iteration requires multiple gradient evaluations and re-optimizations over all accumulated parameters.

\subsection{Graph Embedding via FEATHER}

To integrate structural information from problem graphs into the generative model, we employ the FEATHER embedding algorithm \cite{rozemberczki2020characteristic}. FEATHER encodes each node $u\in V$ through a characteristic function based on random-walk transition probabilities:
\be
    \phi_u(\theta, r)=\sum_{w \in V} \hat{A}_{u, w}^r  e^{i \theta x_w} \ ,
\ee
where  $\hat{A} = D^{-1}A$ is the normalized adjacency matrix, $r$ is the walk length, and $x_w$ represents node attributes.

Averaging $\phi_u$ across sampled frequencies $\theta$ yields a compact, complex-valued representation of the node’s higher-order connectivity pattern.

The resulting node-level embeddings are mean-pooled to obtain an isomorphism-invariant graph descriptor. FEATHER embeddings thus provide the generative model with continuous structural features that capture graph topology and local correlation structure, enabling it to generalize across diverse problem instances.

\subsection{Generative Model: QAOA-GPT}

We utilize a modified version of nanoGPT \cite{Karpathy}, an implementation of the decoder-only Transformer architecture \cite{NIPS2017_3f5ee243}, following the QAOA-GPT framework of Tyagin \textit{et al.}\ \cite{tyagin2025qaoa}, but extend it to handle HUBO Hamiltonians that include linear, quadratic, and cubic interaction terms. The model is trained on paired datasets of graphs and corresponding ADAPT-QAOA circuits, treating circuit synthesis as a sequence modeling task. Each training sample encodes a spin-glass instance and its optimized circuit as a token sequence comprising graph descriptors and layerwise circuit parameters.

Each instance is represented by two token streams:
\begin{itemize}
    \item Graph tokens, which encode the coefficients $d_v, d_{ij}, d_{ijk} \in \{ -1,+1 \}$ of the cost Hamiltonian,
    \item Circuit tokens, which specify the operator indices and optimized parameters $(o_k, \beta_k, \gamma_k)$ for each circuit layer.
\end{itemize}
Numerical parameters are rounded to two decimal places and clipped to $[-10,10]$. Graph embeddings computed via FEATHER are broadcast across all tokens, providing global context.

The model is trained autoregressively to minimize next-token prediction loss, enabling it to generate full circuits conditioned on graph embeddings. Early stopping is guided by two metrics: \textit{(i)} the average approximation ratio on a validation set, and \textit{(ii)} the structural validity of generated circuits.

\subsection{Higher-Order Spin-Glass Hamiltonians}

The HUBO problem instances used for training and evaluation are derived from spin-glass Hamiltonians of the form
\begin{equation}
    H_C =
    \sum_{v \in V} d_v Z_v
    + \sum_{(i,j) \in E} d_{ij} Z_i Z_j  + \sum_{(i,j,k) \in W} d_{ijk} Z_i Z_j Z_k
\end{equation}
where $d_v, d_{ij}, d_{ijk}$ are random coupling strengths chosen from $\{ -1,+1 \}$.

The inclusion of cubic terms introduces higher-order frustration, producing rugged energy landscapes that make these models suitable benchmarks for quantum optimization algorithms.

\begin{figure}[htb]
    \centering
    \begin{subfigure}{0.48\textwidth}
        \centering
        \includegraphics[width=\linewidth]{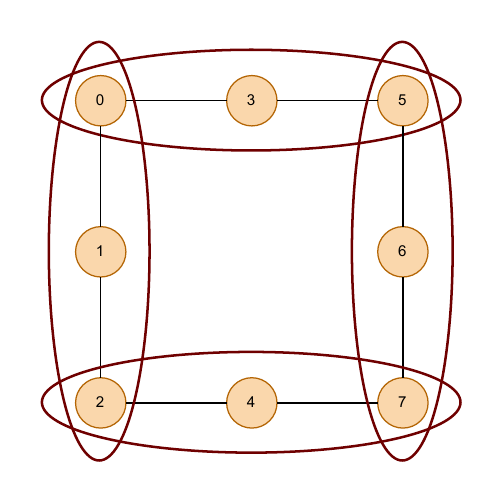}
        \caption{8-qubit graph}
        \label{fig:sg_8n}
    \end{subfigure}
    \\
    \begin{subfigure}{0.48\textwidth}
        \centering
        \includegraphics[width=\linewidth]{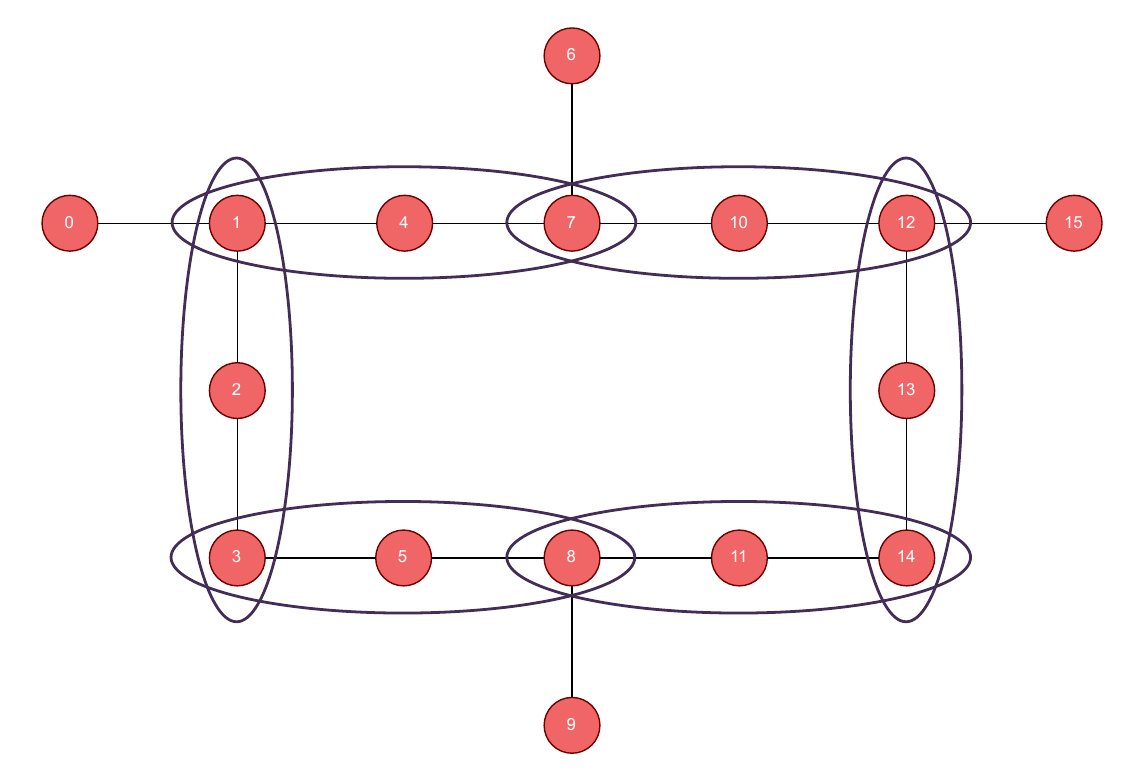}
        \caption{16-qubit graph}
        \label{fig:sg_16n}
    \end{subfigure}
    \caption{Heavy-hex graph topologies used for spin glass problem generation. (a) 8-qubit graph. (b) 16-qubit graph. Nodes represent qubits (linear terms $Z_v$ on nodes in $V$), edges represent quadratic interactions ($Z_iZ_j$) along edges in $E$, and hyperedges in $W$ encircling three neighboring nodes represent cubic interaction terms ($Z_iZ_jZ_k$) of the HUBO Hamiltonian. All Ising coefficients $d_v, d_{ij}, d_{ijk}$ were uniformly sampled from $\{-1,+1\}$.}
    \label{fig:sg_grids}
\end{figure}

The training dataset comprises 8- and 16-qubit instances embedded on heavy-hex topologies (depicted in Figure \ref{fig:sg_grids}) to reflect realistic hardware connectivity constraints. Circuits achieving a target approximation ratio $\alpha \ge 0.87$ are retained for training to ensure high-quality supervision.

This methodological framework allows the model to learn mappings from graph-encoded Hamiltonians to optimized circuit representations. All quantum circuit simulations for generating training data and evaluating generated circuits were performed using the CUDA-Q platform \cite{nvidia_cudaq}, leveraging its high-performance simulator on classical HPC resources. Once trained, QAOA-GPT can generate adaptive QAOA-like circuits for unseen HUBO instances in a single forward pass, effectively bypassing classical parameter optimization.

\subsection{Performance Metric: Approximation Ratio}

To quantify the quality of solutions produced by either ADAPT-QAOA or the generative model, we employ the approximation ratio
\begin{equation}
\alpha = \frac{\langle \psi(\boldsymbol{\gamma}, \boldsymbol{\beta}) | H_C | \psi(\boldsymbol{\gamma}, \boldsymbol{\beta}) \rangle}{E_\text{OPT}(G)}
\end{equation}
where $H_C$ is the cost Hamiltonian associated with the problem instance and $E_\text{OPT}(G)$ denotes its exact ground-state energy obtained from a classical solver.

Because both $\braket{H_C}$ and $E_\text{OPT}(G)$ are negative for minimization problems such as spin-glass Hamiltonians, we take their absolute values when reporting ratios so that $0 < \alpha \le 1$, with $\alpha =1$ corresponding to the exact ground state.
This normalization ensures a consistent comparison across instances of different sizes and coupling distributions.

Throughout this work, we report both the best and average approximation ratios over test-set instances to evaluate the model’s overall accuracy and generalization.

The methodological framework described above enables QAOA-GPT to map higher-order spin-glass Hamiltonians directly to optimized variational circuits.

In the following Section, we present numerical results demonstrating the model’s performance on 8- and 16-qubit HUBO instances, analyze the learned parameter distributions, and compare the generated circuits with those obtained from classical ADAPT-QAOA optimization.

\begin{figure*}[htb]
    \includegraphics[width=0.9\textwidth]{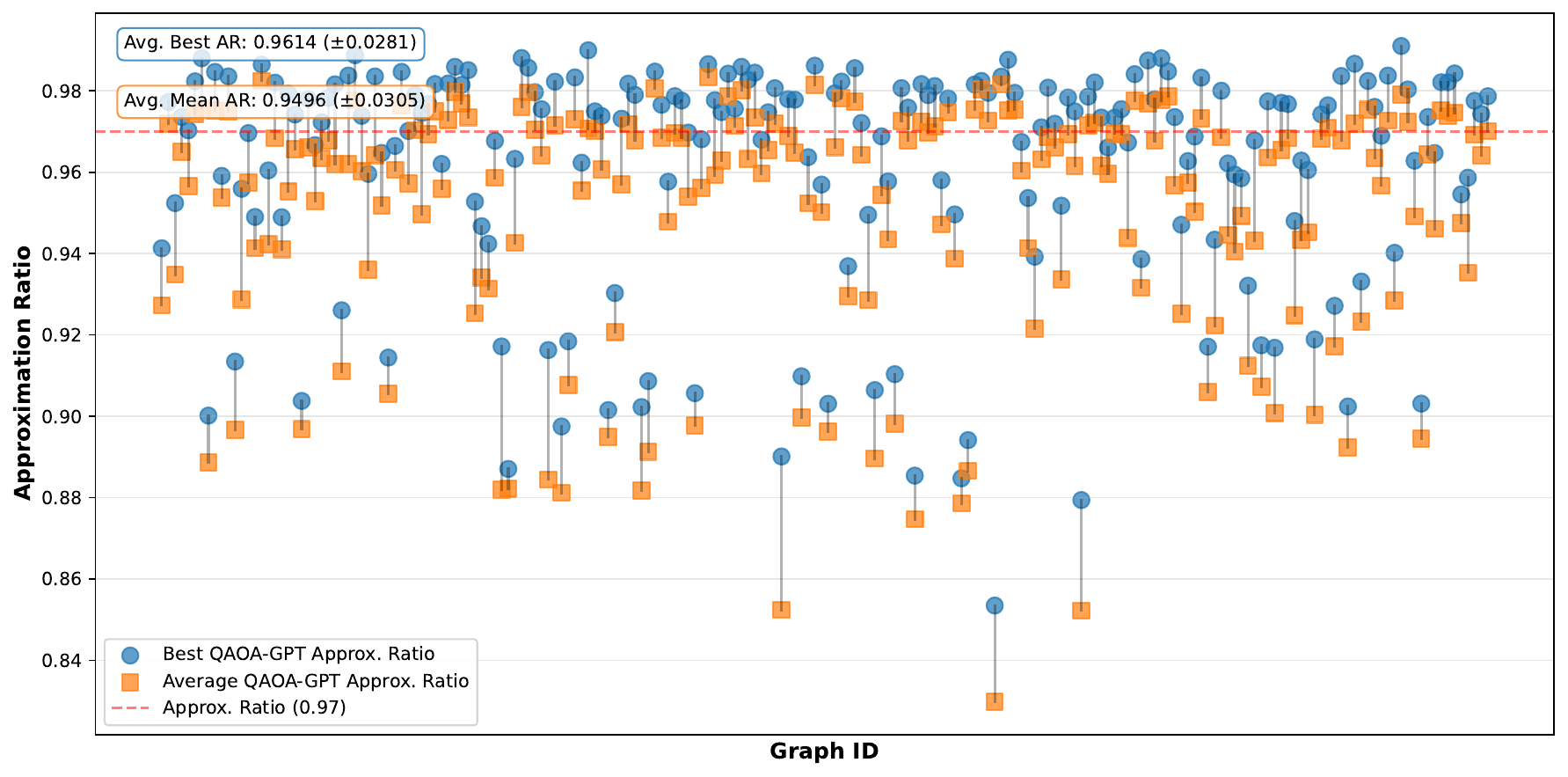}
    \caption{Performance distribution of QAOA-GPT on 200 16-qubit spin glass instances at depth \( p_{\text{max}} = 15 \). For each instance, the model generated 10 circuits. The `Best QAOA-GPT Approx. Ratio` (blue) shows the highest approximation ratio (\( \alpha \)) achieved per instance, averaging \( 0.9614 \pm 0.0281 \). The `Average QAOA-GPT Approx. Ratio` (orange) shows the mean performance across all 10 generated circuits per instance, averaging \( 0.9496 \pm 0.0305 \). The dashed line indicates the target approximation ratio of 0.97.}
    \label{fig:16n_ar}
\end{figure*}

\begin{figure*}[htb]
    \includegraphics[width=0.9\textwidth]{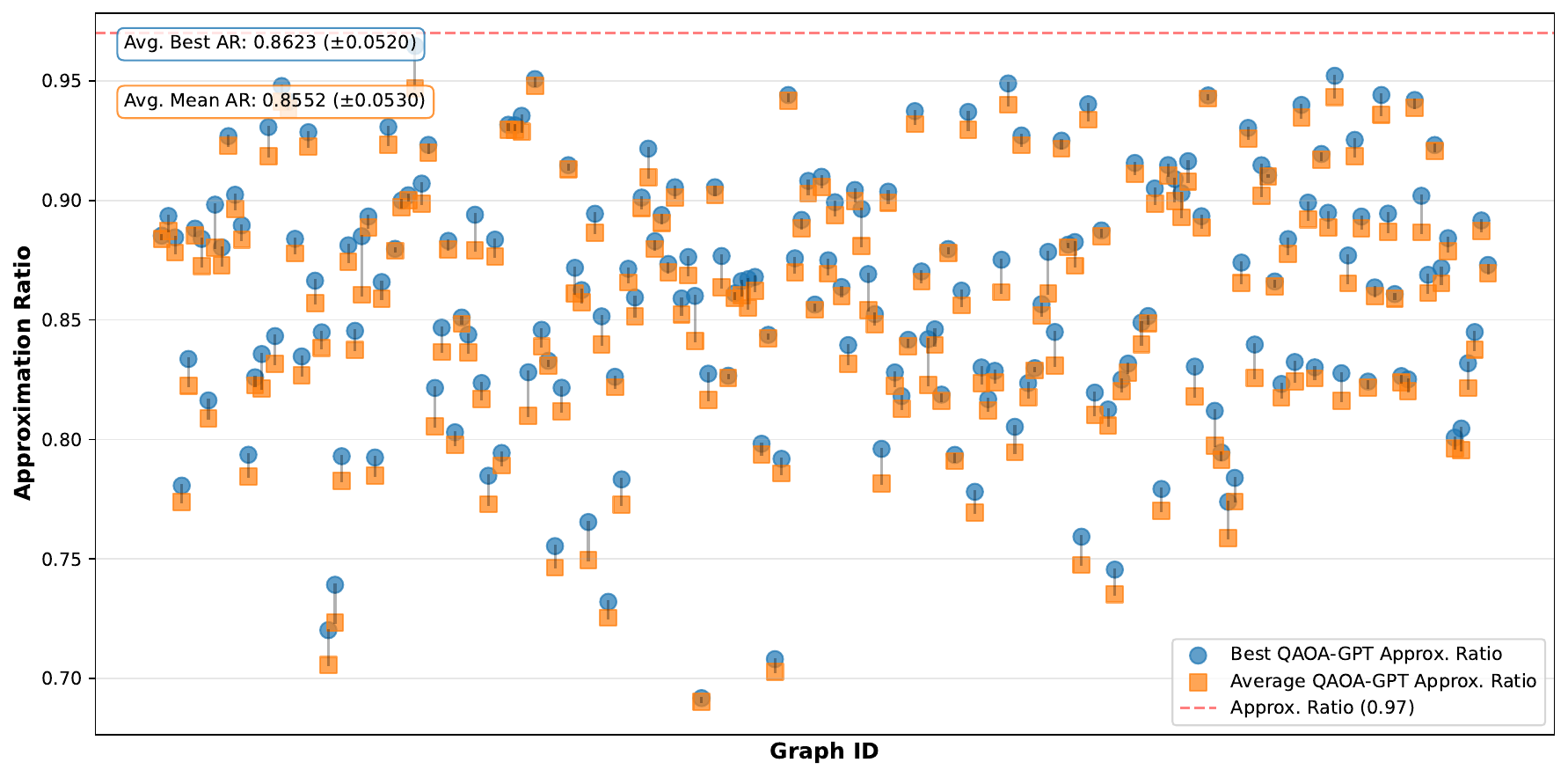}
    \caption{Performance distribution of QAOA-GPT on 200 16-qubit spin glass instances at constrained depth \( p_{\text{max}} = 5 \). For each instance, the model generated 10 circuits. The `Best QAOA-GPT Approx. Ratio` (blue) shows the highest approximation ratio (\( \alpha \)) achieved per instance, averaging \( 0.8623 \pm 0.0520 \). The `Average QAOA-GPT Approx. Ratio` (orange) shows the mean performance across all 10 generated circuits per instance, averaging \( 0.8552 \pm 0.0530 \). The dashed line indicates the target approximation ratio of 0.97.}
    \label{fig:16n_5l_ar}
\end{figure*}

\section{Results and Discussion}  \label{sec:results}

We evaluate the performance of the extended QAOA-GPT model on ensembles of 8- and 16-qubit higher-order spin-glass instances defined on heavy-hex graphs. Each instance contains linear, quadratic, and cubic coupling terms with coefficients $d_v, d_{ij}, d_{ijk} \in \{-1, +1\}$. For both system sizes, the dataset was partitioned into non-overlapping training and test subsets, and only ADAPT-QAOA circuits achieving a minimum approximation ratio $\alpha \ge 0.87$ were used for model training. The trained model was then evaluated on unseen instances to assess its capacity for generalization and autonomous circuit generation.

\subsection{Overall Performance}

Figures \ref{fig:16n_ar} and \ref{fig:16n_5l_ar} summarize the approximation ratios obtained for 16-qubit test instances for $p_\text{max} =15$ and $p_\text{max} =5$, respectively.
The generative model achieves average and best-case approximation ratios of $0.9496 \pm 0.0305$ and $ 0.9614 \pm 0.0281$, respectively, closely matching the results obtained from classically optimized ADAPT-QAOA circuits used in training. For a summary, see Table \ref{tab:16n_results}.
\begin{table}[htb]
\centering
\caption{Approximation Ratios for 16-Qubit QAOA-GPT Circuits Across Depth Configurations}
\begin{tabular}{c@{\hspace{2em}}c@{\hspace{2em}}c}
\hline
Circuit Depth ($p_{\text{max}}$) & Mean AR & Best AR \\
\hline
15 & $0.950 \pm 0.031$ & $0.961 \pm 0.028$ \\
5 & $0.855 \pm 0.053$ & $0.862 \pm 0.052$ \\
\hline
\end{tabular}
\label{tab:16n_results}
\end{table}

\begin{figure*}[htb]
    \includegraphics[width=0.9\textwidth]{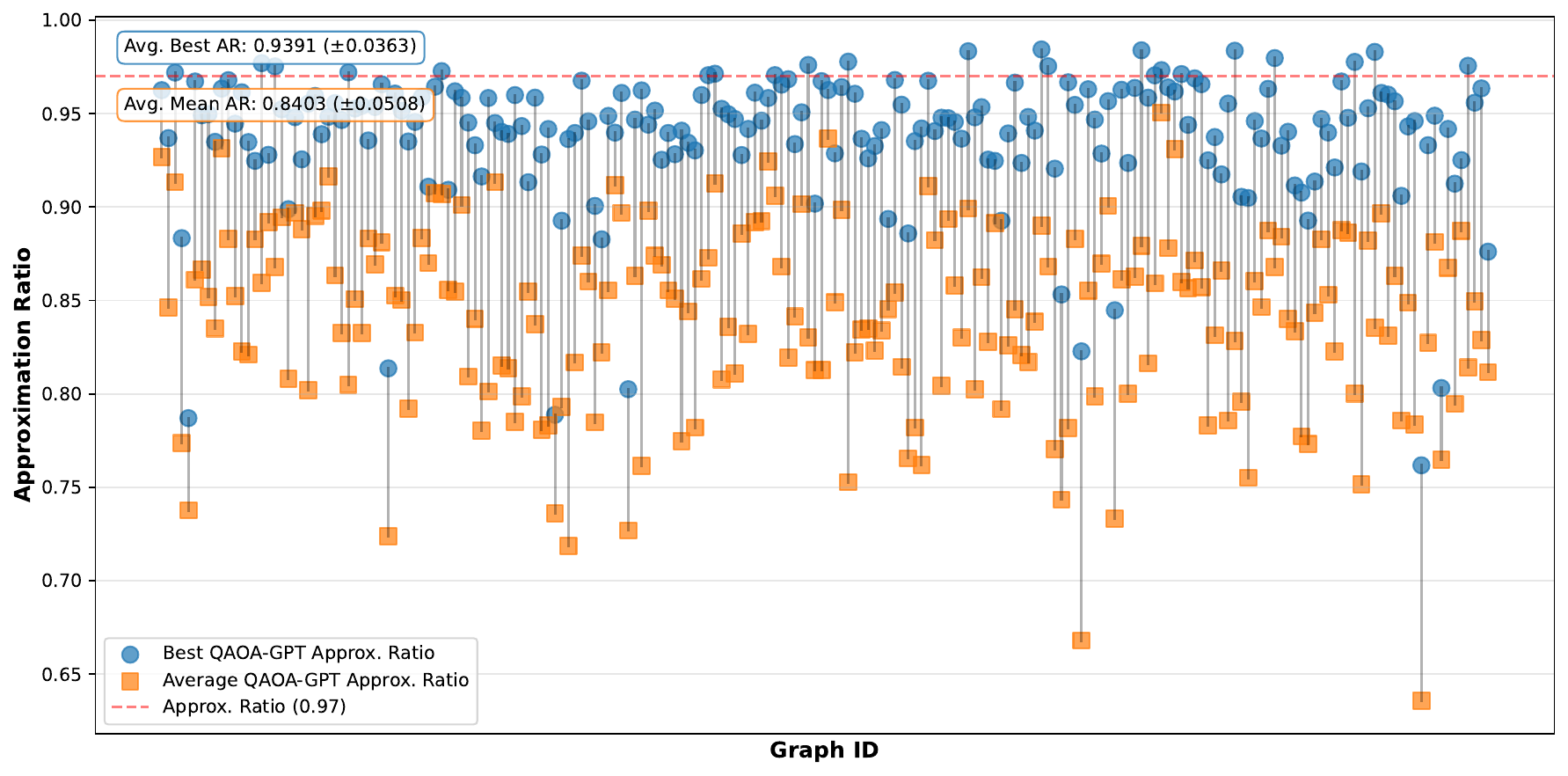}
    \caption{Performance distribution of QAOA-GPT on 200 8-qubit spin glass instances at depth \( p_{\text{max}} = 15 \). For each instance, the model generated 10 circuits. The `Best QAOA-GPT Approx. Ratio` (blue) shows the highest approximation ratio (\( \alpha \)) achieved per instance, averaging \( 0.9391 \pm 0.0363 \). The `Average QAOA-GPT Approx. Ratio` (orange) shows the mean performance across all 10 generated circuits per instance, averaging \( 0.8403 \pm 0.0508 \). The dashed line indicates the target approximation ratio of 0.97.}
    \label{fig:8n_ar}
\end{figure*}

\begin{figure*}[htb]
    \includegraphics[width=0.9\textwidth]{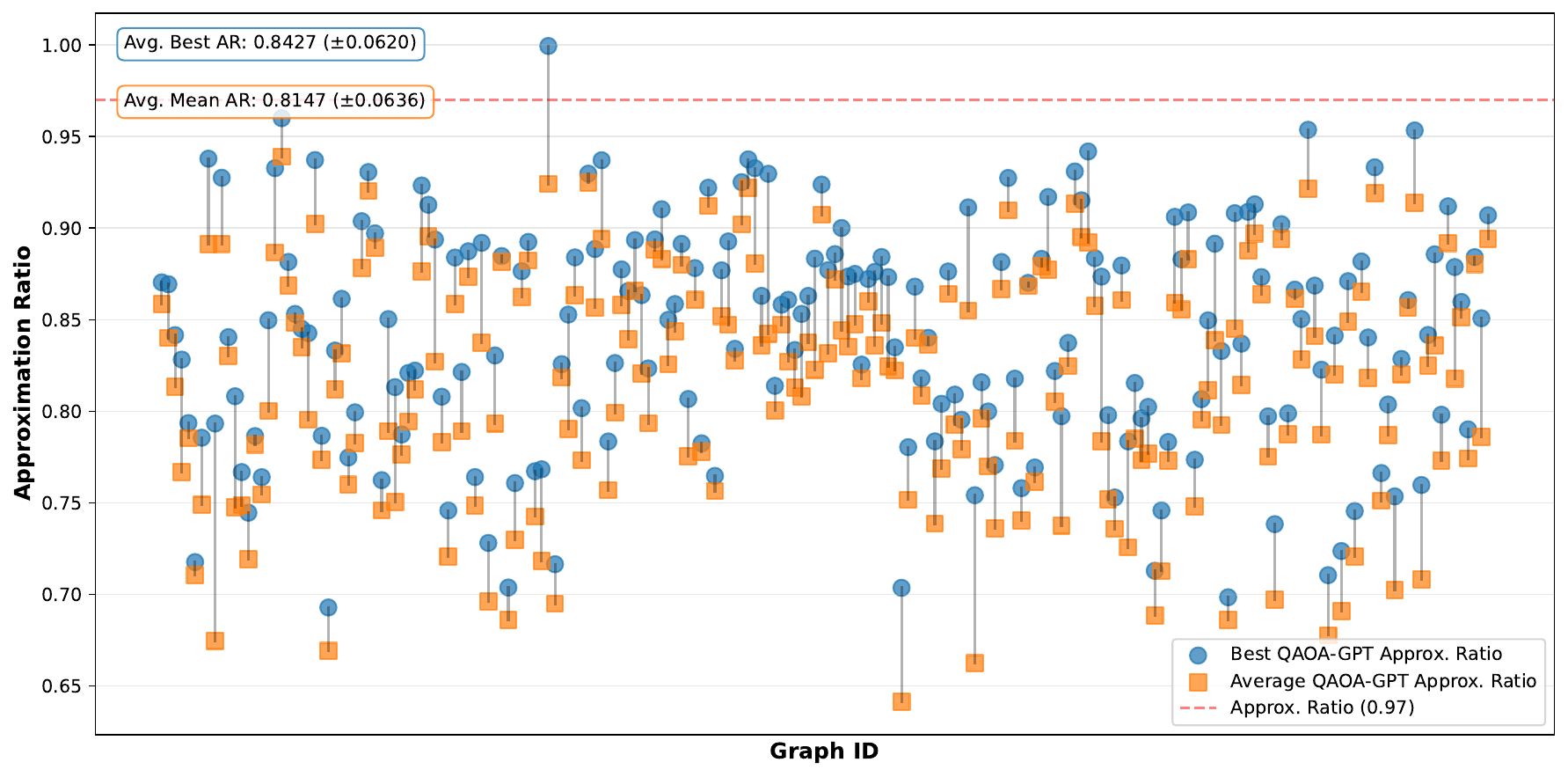}
    \caption{Performance distribution of QAOA-GPT on 200 8-qubit spin glass instances at constrained depth \( p_{\text{max}} = 5 \). For each instance, the model generated 10 circuits. The `Best QAOA-GPT Approx. Ratio` (blue) shows the highest approximation ratio (\( \alpha \)) achieved per instance, averaging \( 0.8427 \pm 0.0620 \). The `Average QAOA-GPT Approx. Ratio` (orange) shows the mean performance across all 10 generated circuits per instance, averaging \( 0.8147 \pm 0.0636 \). The dashed line indicates the target approximation ratio of 0.97.}
    \label{fig:8n_5l_ar}
\end{figure*}

For 8-qubit systems, the corresponding mean ratio is approximately 0.93. Details are depicted in Figure \ref{fig:8n_ar} for  $p_\text{max} =15$ and Figure \ref{fig:8n_5l_ar} for $p_\text{max} =5$. For a summary, see Table \ref{tab:8n_results}.
\begin{table}[htb]
\centering
\caption{Approximation Ratios for 8-Qubit QAOA-GPT Circuits Across Depth Configurations}
\begin{tabular}{c@{\hspace{2em}}c@{\hspace{2em}}c@{\hspace{1em}}}
\hline
Circuit Depth ($p_{\text{\text{max}}}$) & Mean AR & Best AR \\
\hline
15 & $0.840 \pm 0.051$ & $0.940 \pm 0.036$ \\
5 & $0.815 \pm 0.064$ & $0.843 \pm 0.062$ \\
\hline
\end{tabular}
\label{tab:8n_results}
\end{table}

These values confirm that QAOA-GPT maintains high solution quality while completely bypassing classical variational optimization.

The agreement between generated and reference circuits demonstrates that the model has learned an effective internal representation of the mapping from graph structure to optimized circuit parameters.
In particular, the consistency of $\alpha$
across system sizes indicates that the generative approach scales favorably with the number of qubits and the order of interaction terms, a non-trivial result given the combinatorial growth of the underlying configuration space.

\subsection{Scaling Behavior and Model Generalization}

The increase in qubit count from 8 to 16 provides an opportunity to examine how well the model generalizes to larger graphs and more complex energy landscapes.
Interestingly, the 16-qubit results exhibit both a higher mean approximation ratio and smaller variance than the 8-qubit case.
This behavior suggests that the transformer’s contextual representation benefits from richer graph embeddings, allowing it to capture higher-order connectivity patterns more effectively.
Moreover, as the training set for 16-qubit instances contains more structurally diverse graphs, the model likely develops a smoother latent representation of the problem–circuit mapping, leading to more stable performance.

The absence of any noticeable degradation in $\alpha$
with increasing system size implies that the generative approach can scale to larger HUBO problems without the exponential runtime penalties characteristic of classical parameter-optimization loops.
In practice, circuit generation requires only a single forward pass through the network, reducing computation time from hours to milliseconds.

\subsection{Learned Variational Parameters}

\begin{figure*}[htb]
    \centering
    \begin{subfigure}{0.48\textwidth}
        \centering
        \includegraphics[width=\linewidth]{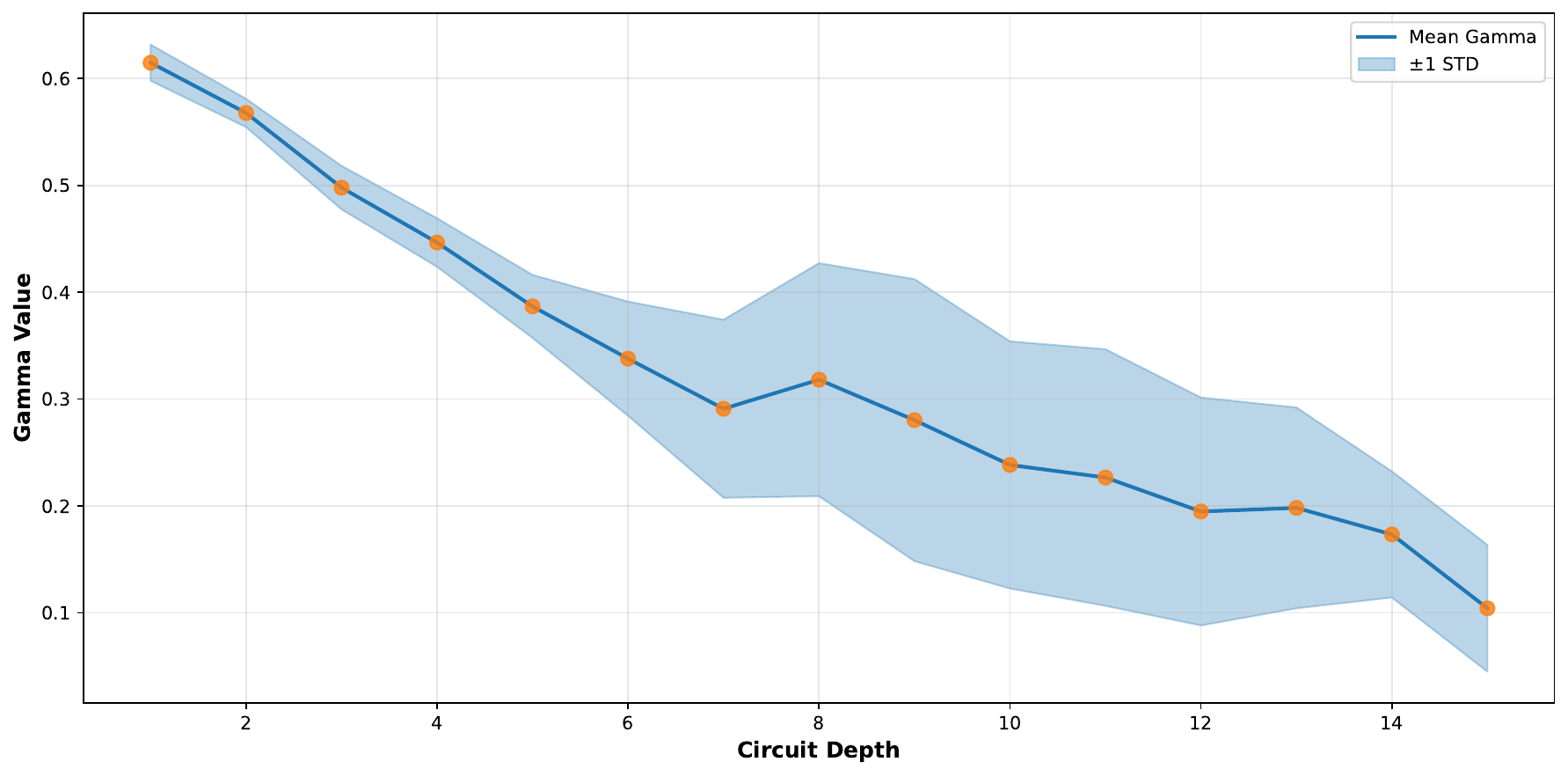}
        \caption{Mean $\gamma$ parameters with standard deviation across circuit depths}
        \label{fig:8n_gscatter}
    \end{subfigure}
    \hfill
    \begin{subfigure}{0.48\textwidth}
        \centering
        \includegraphics[width=\linewidth]{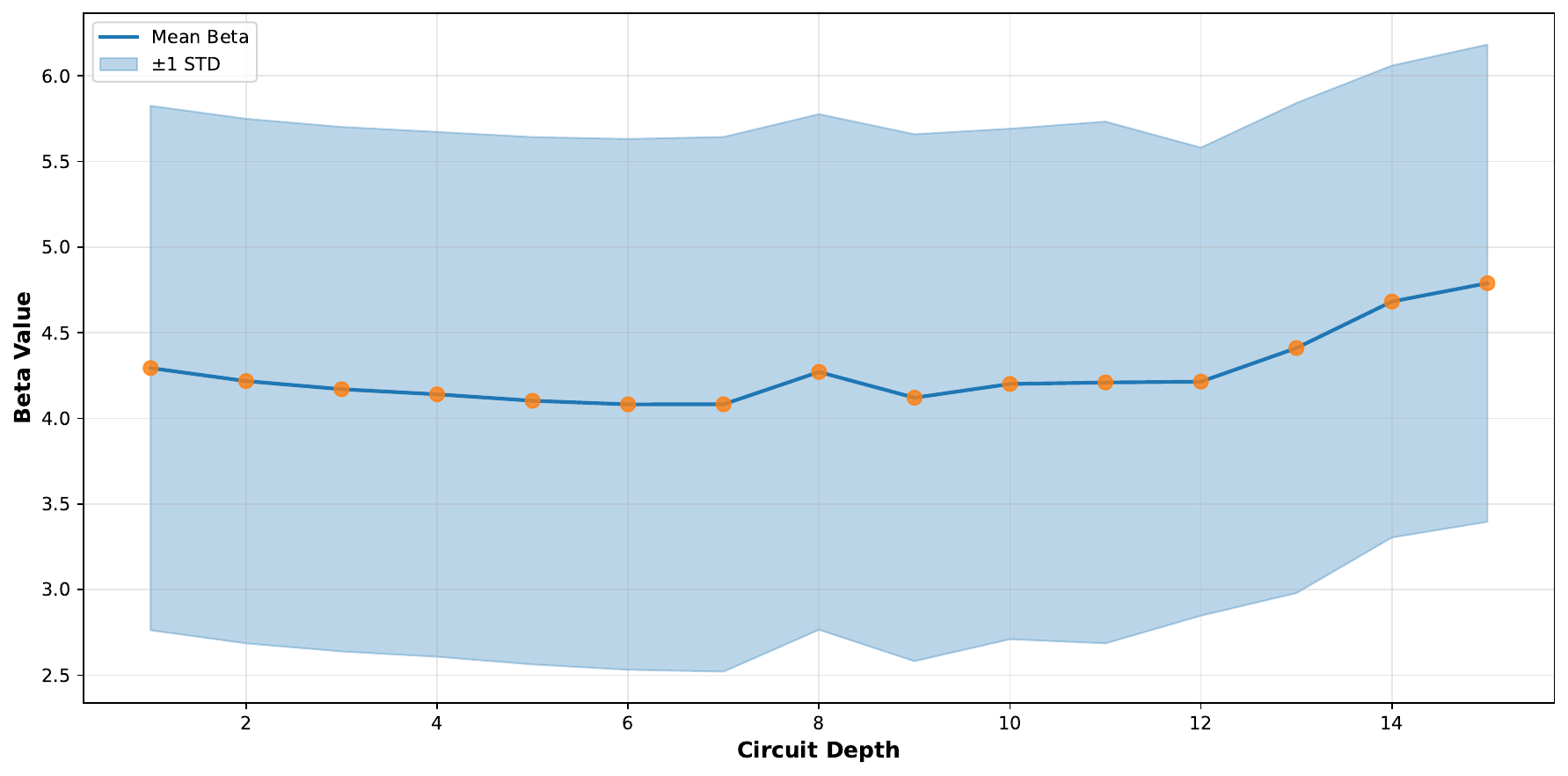}
        \caption{Mean $\beta$ parameters with standard deviation across circuit depths}
        \label{fig:8n_bscatter}
    \end{subfigure}
    \\
    \begin{subfigure}{0.48\textwidth}
        \centering
        \includegraphics[width=\linewidth]{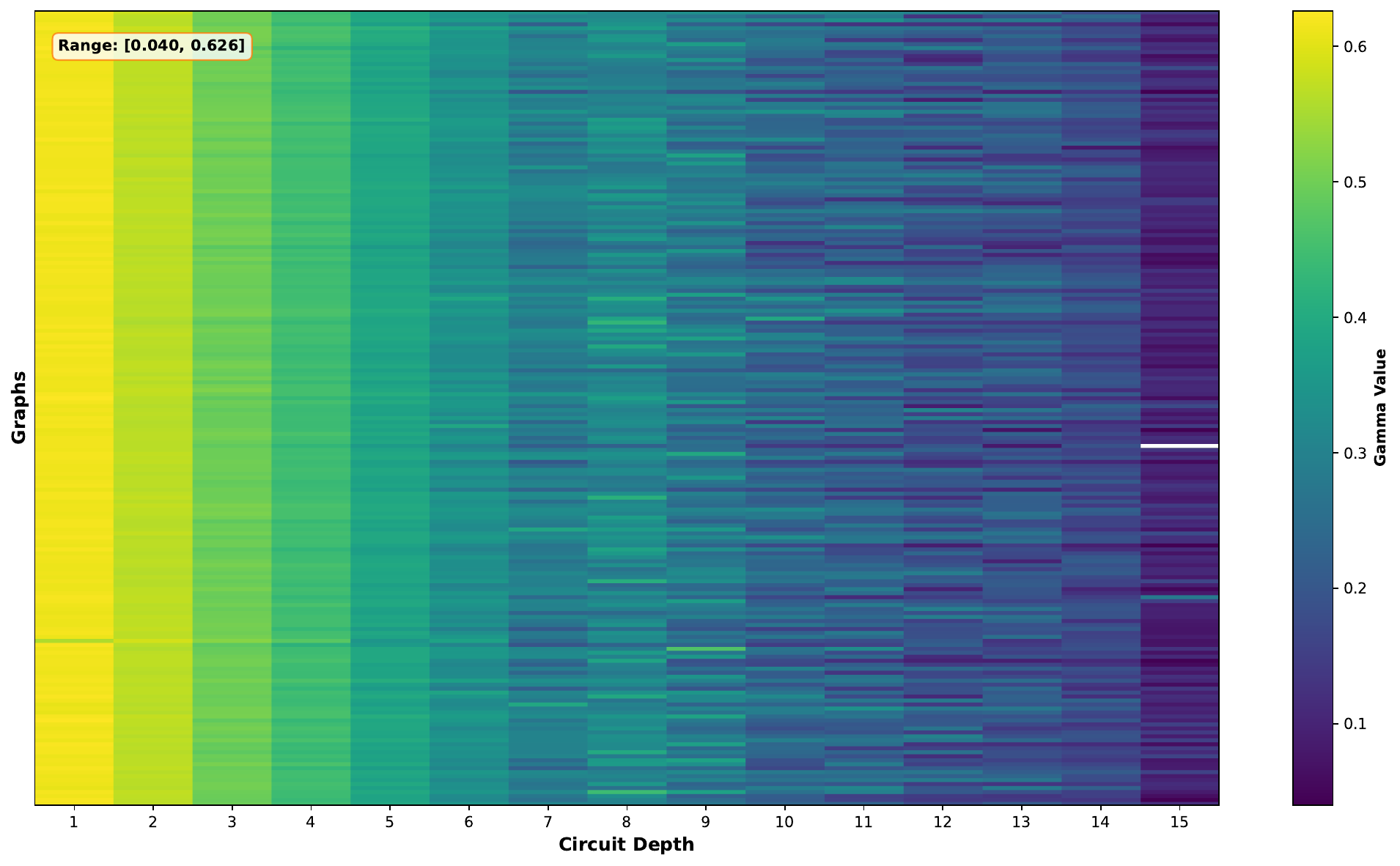}
        \caption{Heatmap of $\gamma$ values across instances and depths}
        \label{fig:8n_gheatmap}
    \end{subfigure}
    \hfill
    \begin{subfigure}{0.48\textwidth}
        \centering
        \includegraphics[width=\linewidth]{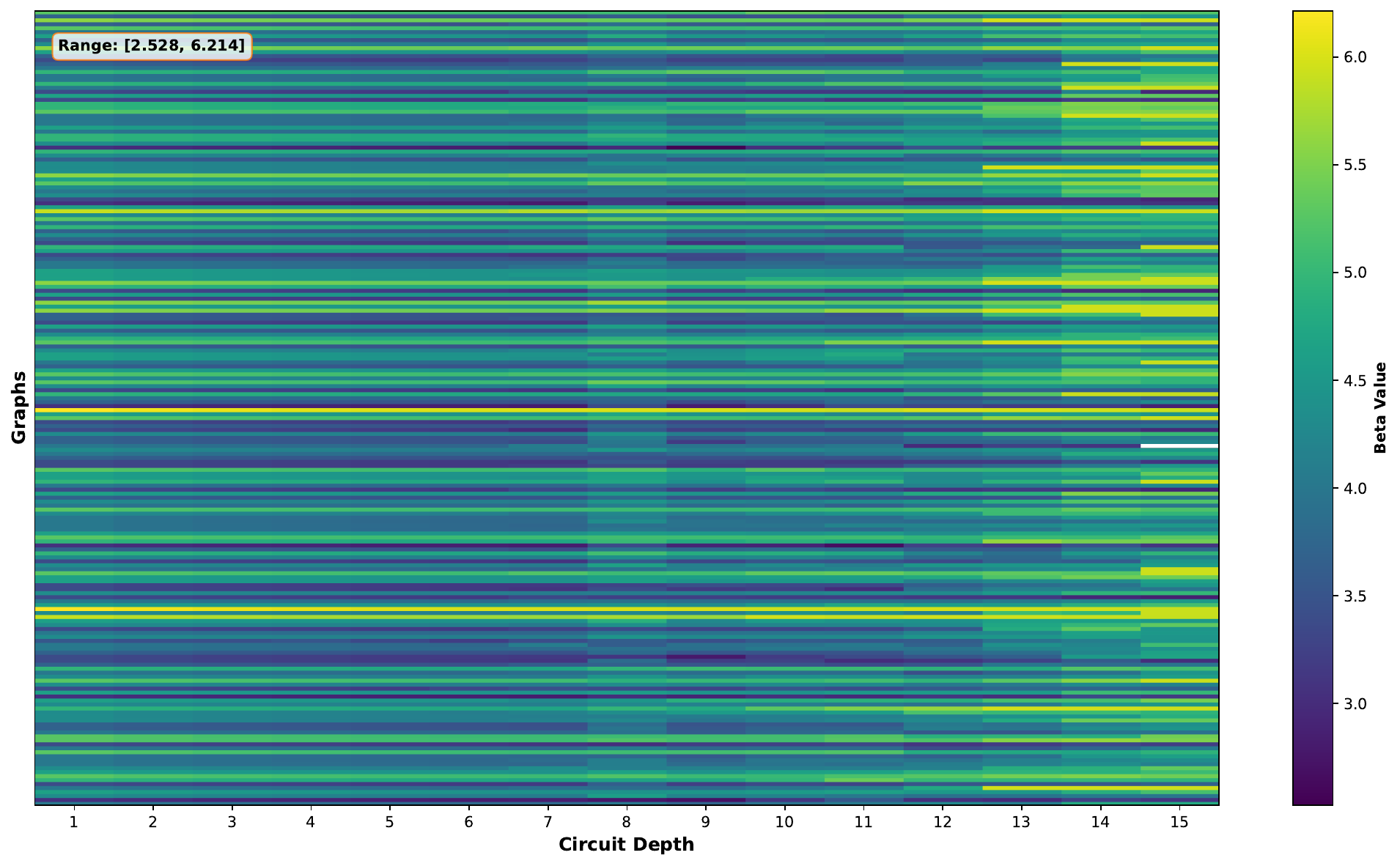}
        \caption{Heatmap of $\beta$ values across instances and depths}
        \label{fig:8n_bheatmap}
    \end{subfigure}
    \caption{Variational parameter distributions for QAOA-GPT generated circuits on 8-qubit spin glass instances at depth \( p_{\text{max}} = 15 \). (a) Mean \( \gamma \) parameters with standard deviation, showing a pronounced decrease across layers. (b) Mean \( \beta \) parameters with standard deviation, maintaining a nearly flat profile. (c) Heatmap of \( \gamma \) values across all graphs and circuit depths, showing concentration in the range [0.040, 0.626]. (d) Heatmap of \( \beta \) values across all graphs and circuit depths, showing values in the range [2.528, 6.214].}
    \label{fig:8n_parameters}
\end{figure*}

\begin{figure*}[htb]
    \centering
    \begin{subfigure}{0.48\textwidth}
        \centering
        \includegraphics[width=\linewidth]{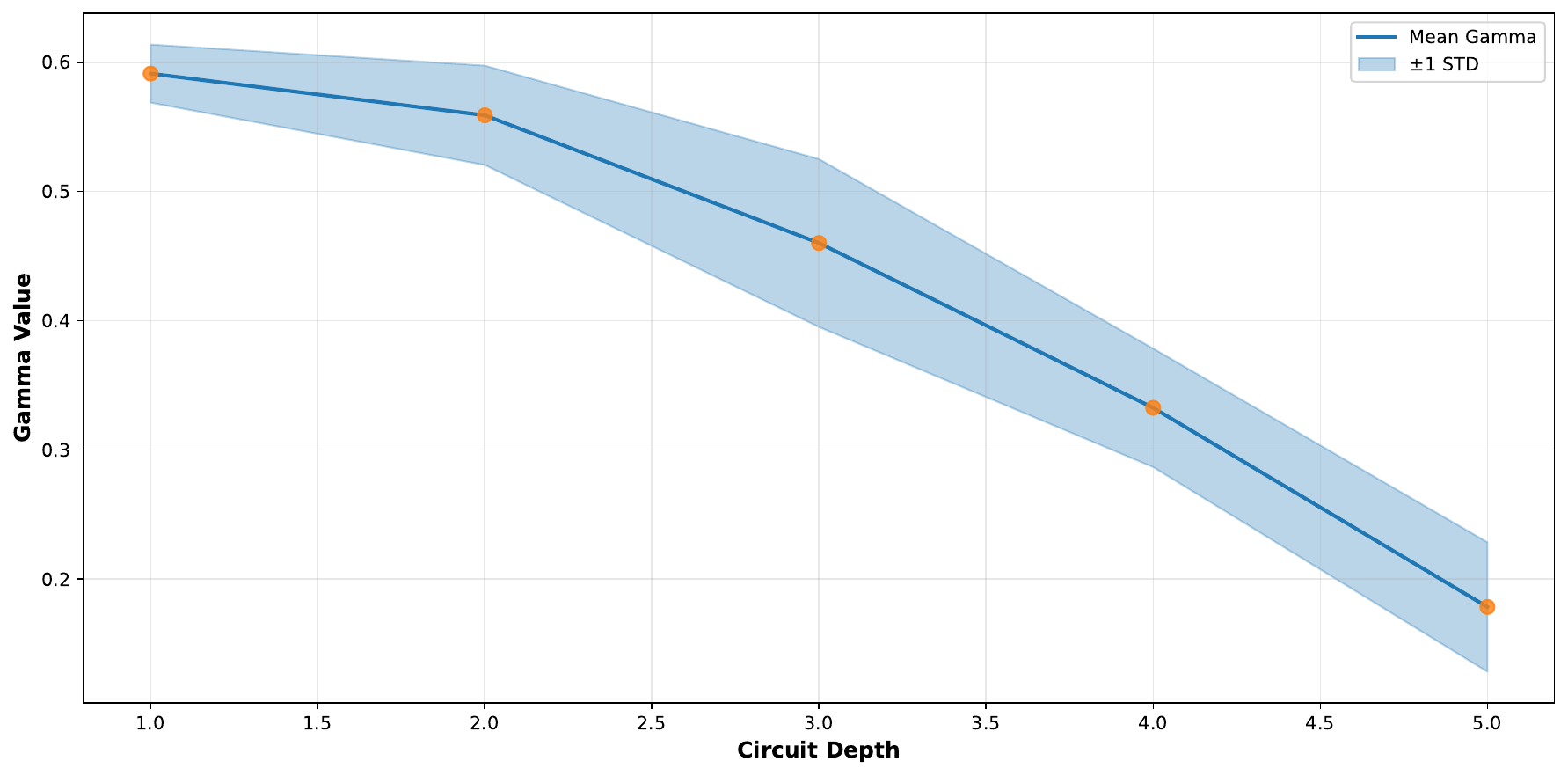}
        \caption{Mean $\gamma$ parameters with standard deviation across circuit depths}
        \label{fig:8n_5l_gscatter}
    \end{subfigure}
    \hfill
    \begin{subfigure}{0.48\textwidth}
        \centering
        \includegraphics[width=\linewidth]{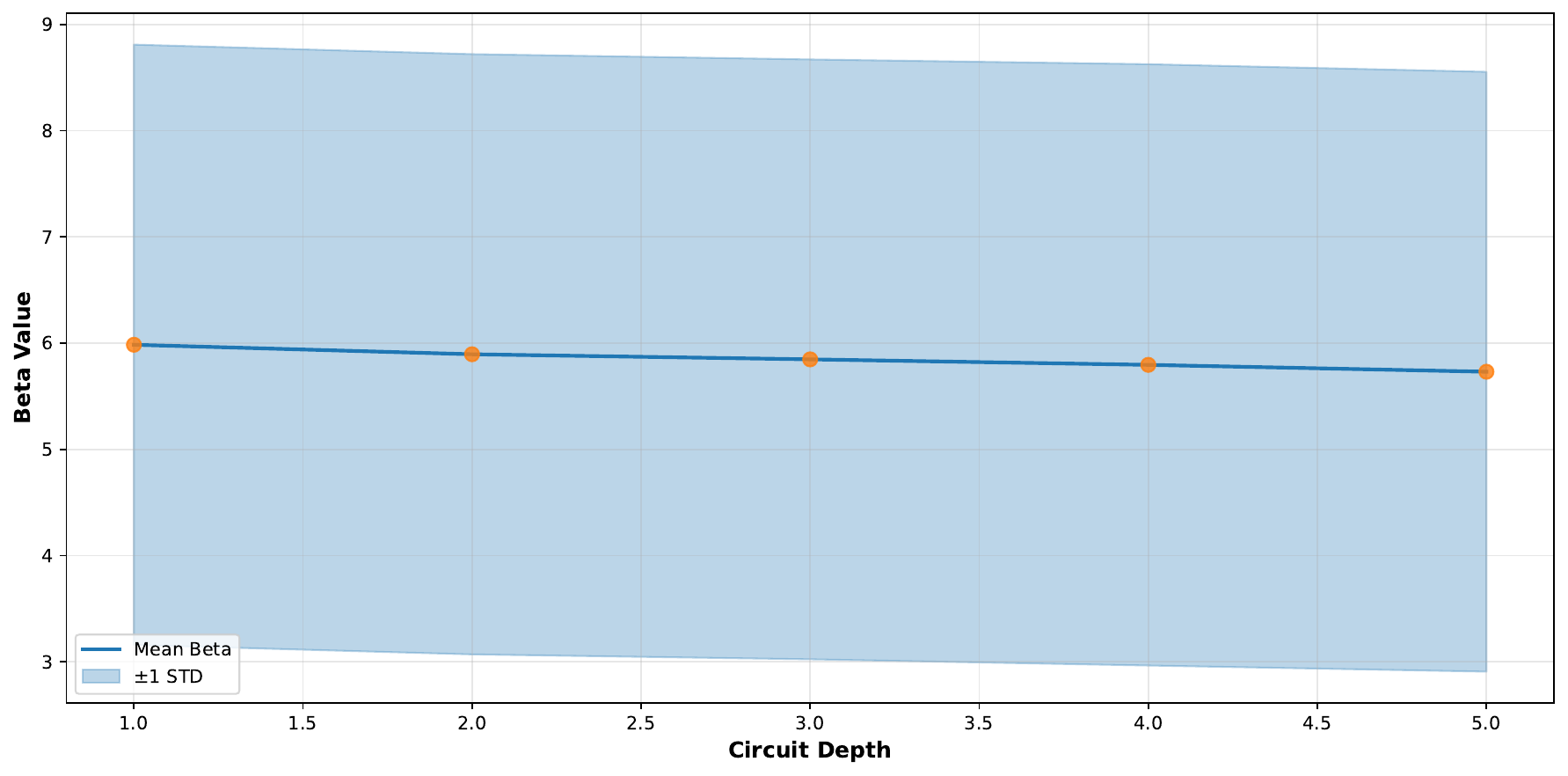}
        \caption{Mean $\beta$ parameters with standard deviation across circuit depths}
        \label{fig:8n_5l_bscatter}
    \end{subfigure}
    \\
    \begin{subfigure}{0.48\textwidth}
        \centering
        \includegraphics[width=\linewidth]{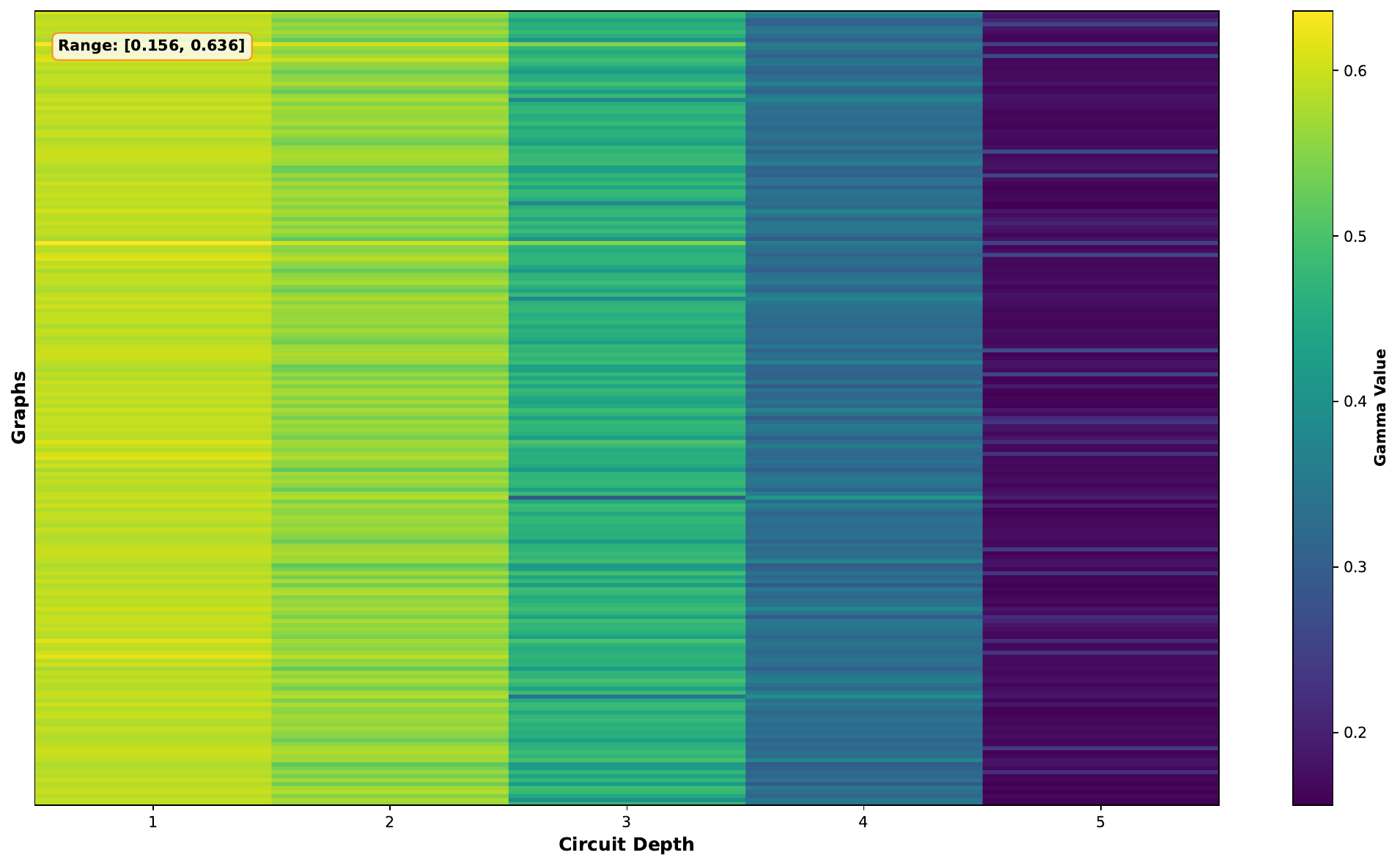}
        \caption{Heatmap of $\gamma$ values across instances and depths}
        \label{fig:8n_5l_gheatmap}
    \end{subfigure}
    \hfill
    \begin{subfigure}{0.48\textwidth}
        \centering
        \includegraphics[width=\linewidth]{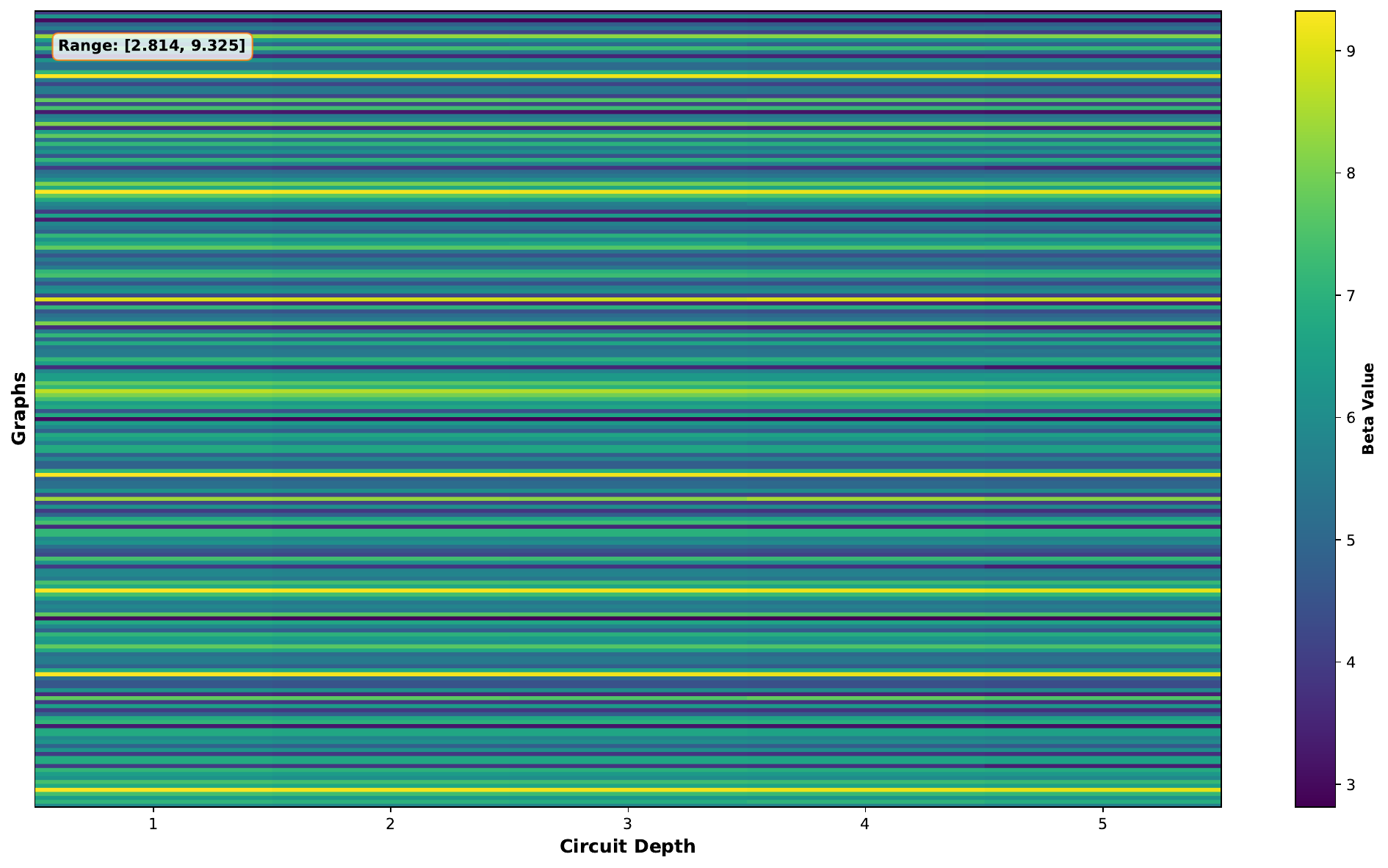}
        \caption{Heatmap of $\beta$ values across instances and depths}
        \label{fig:8n_5l_bheatmap}
    \end{subfigure}
    \caption{Variational parameter distributions for QAOA-GPT generated circuits on 8-qubit spin glass instances at depth \( p_{\text{max}} = 5 \). (a) Mean \( \gamma \) parameters with standard deviation, showing a gentle decrease across layers. (b) Mean \( \beta \) parameters with standard deviation, maintaining a nearly flat profile. (c) Heatmap of \( \gamma \) values across all graphs and circuit depths, showing concentration in the range [0.156, 0.636]. (d) Heatmap of \( \beta \) values across all graphs and circuit depths, showing values in the range [2.814, 9.325].}
    \label{fig:8n_5l_parameters}
\end{figure*}

\begin{figure*}[htb]
    \centering
    \begin{subfigure}{0.48\textwidth}
        \centering
        \includegraphics[width=\linewidth]{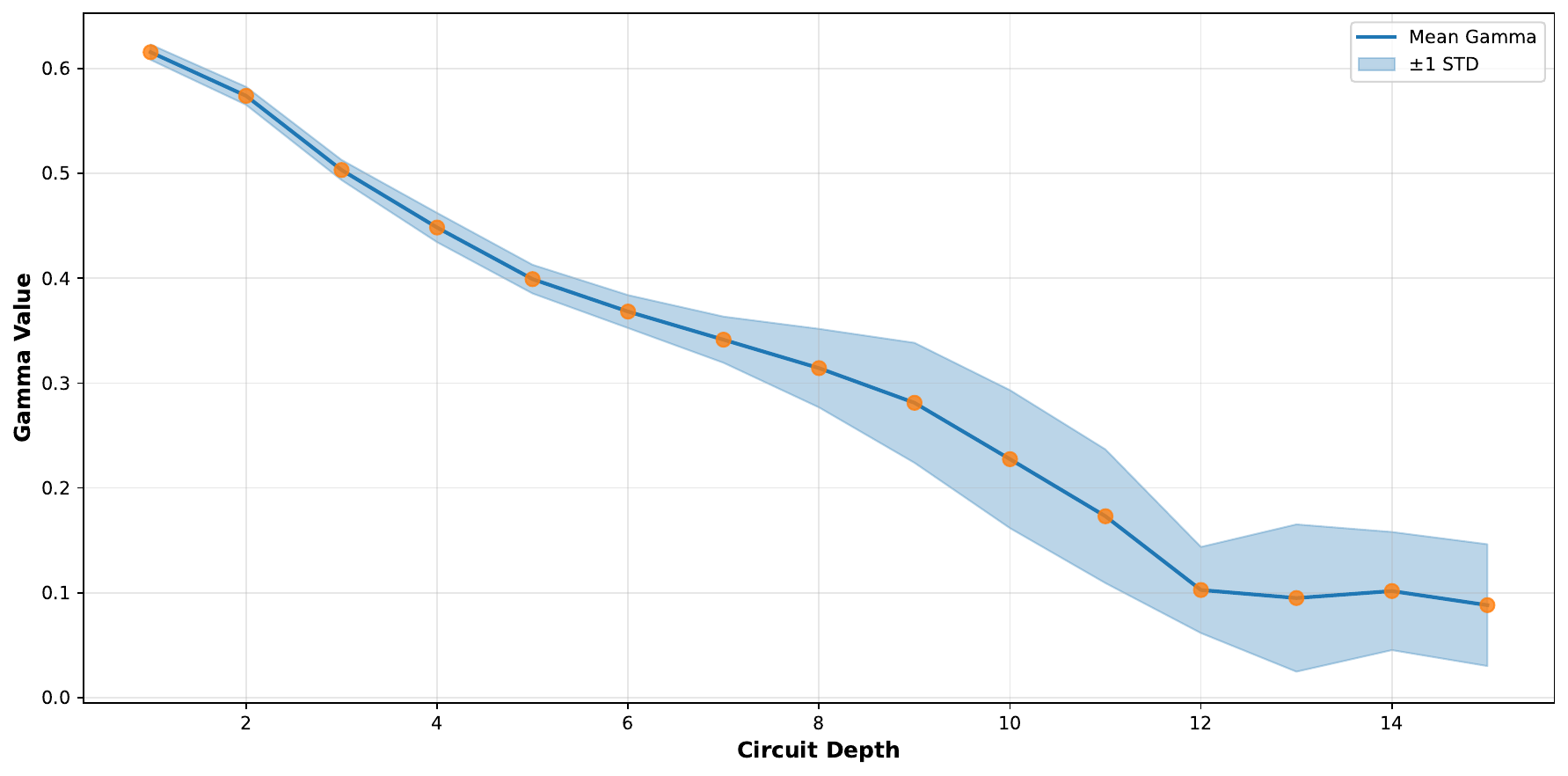}
        \caption{Mean $\gamma$ parameters with standard deviation across circuit depths}
        \label{fig:16n_gscatter}
    \end{subfigure}
    \hfill
    \begin{subfigure}{0.48\textwidth}
        \centering
        \includegraphics[width=\linewidth]{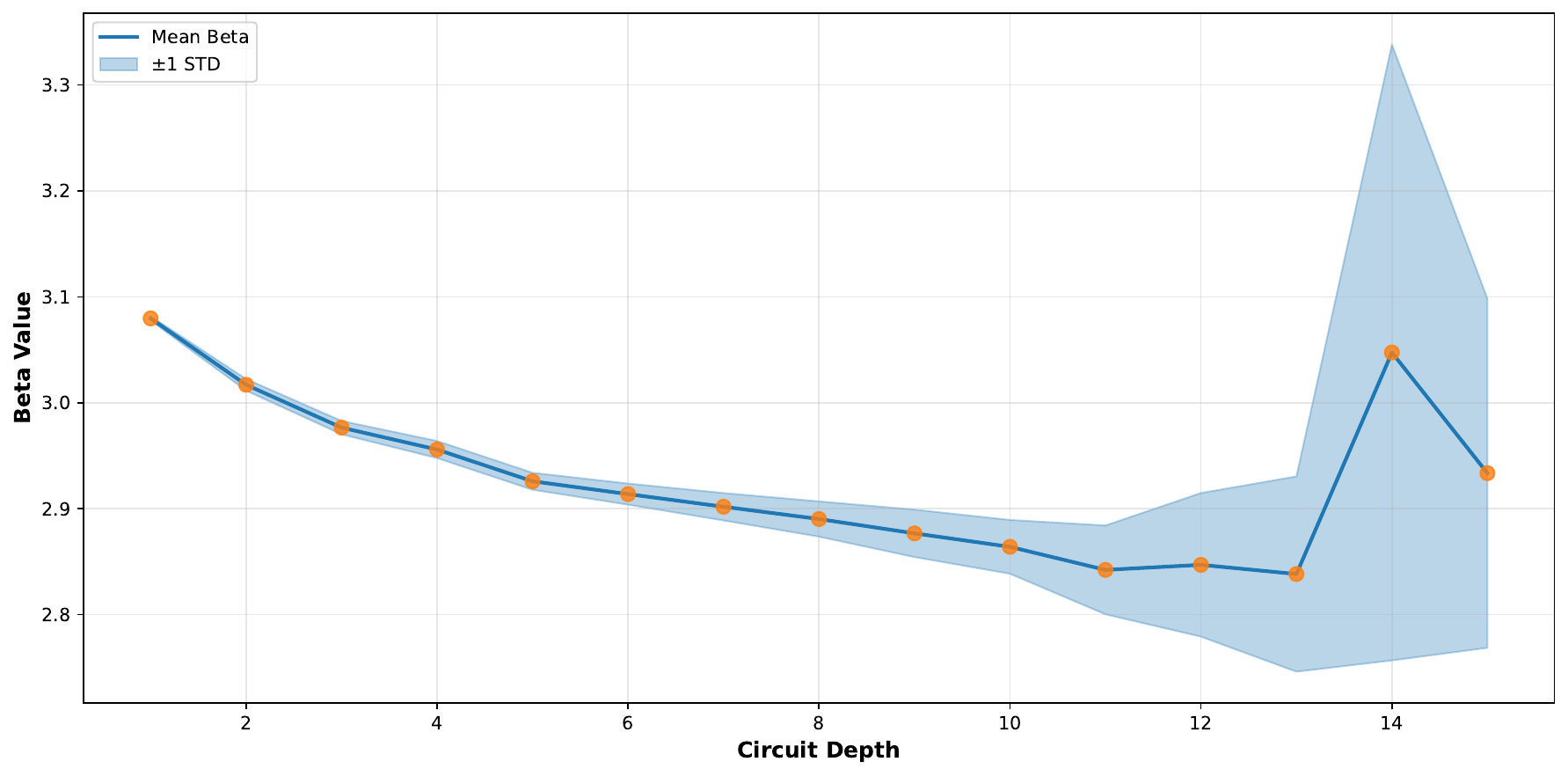}
        \caption{Mean $\beta$ parameters with standard deviation across circuit depths}
        \label{fig:16n_bscatter}
    \end{subfigure}
    \\
    \begin{subfigure}{0.48\textwidth}
        \centering
        \includegraphics[width=\linewidth]{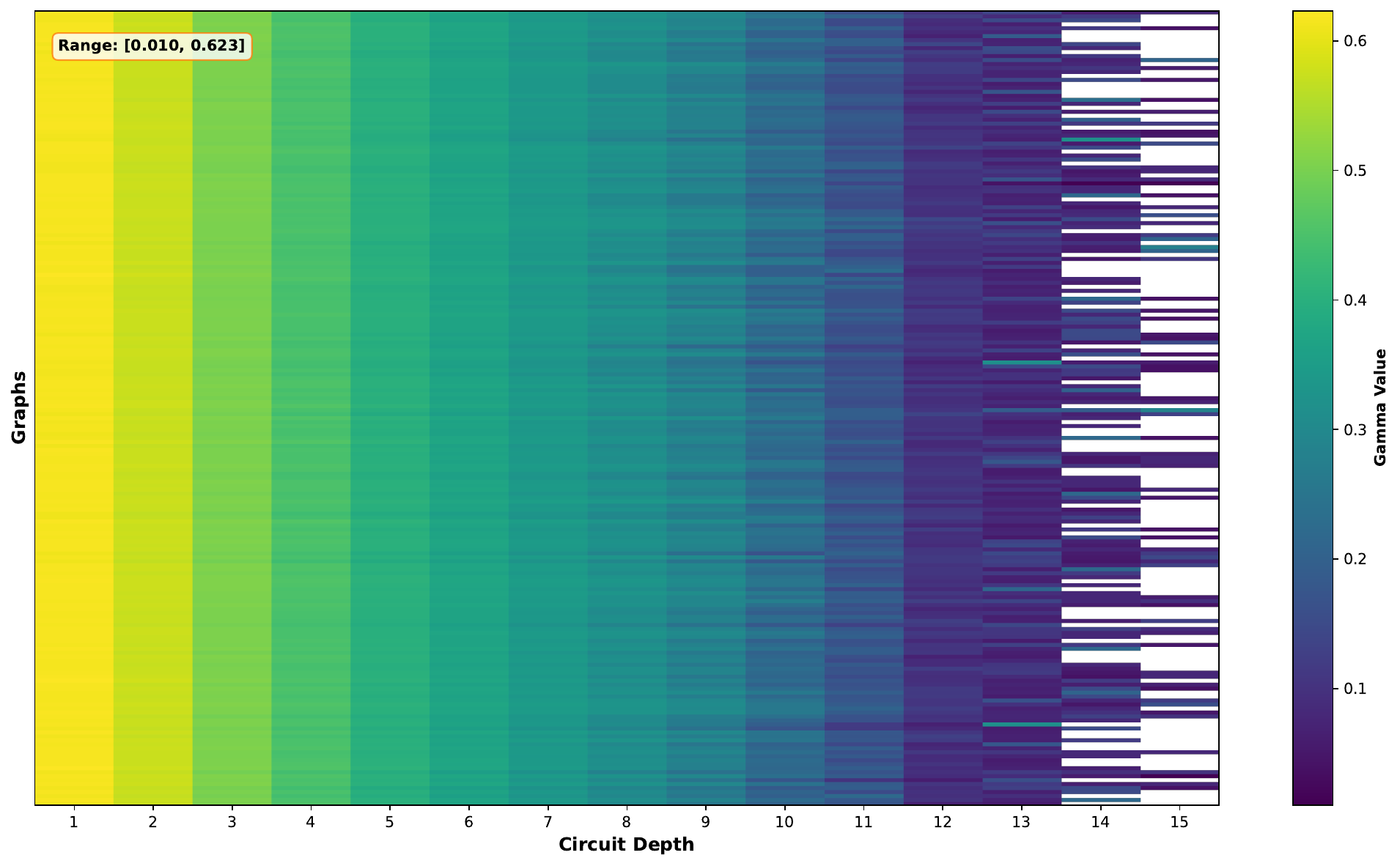}
        \caption{Heatmap of $\gamma$ values across instances and depths}
        \label{fig:16n_gheatmap}
    \end{subfigure}
    \hfill
    \begin{subfigure}{0.48\textwidth}
        \centering
        \includegraphics[width=\linewidth]{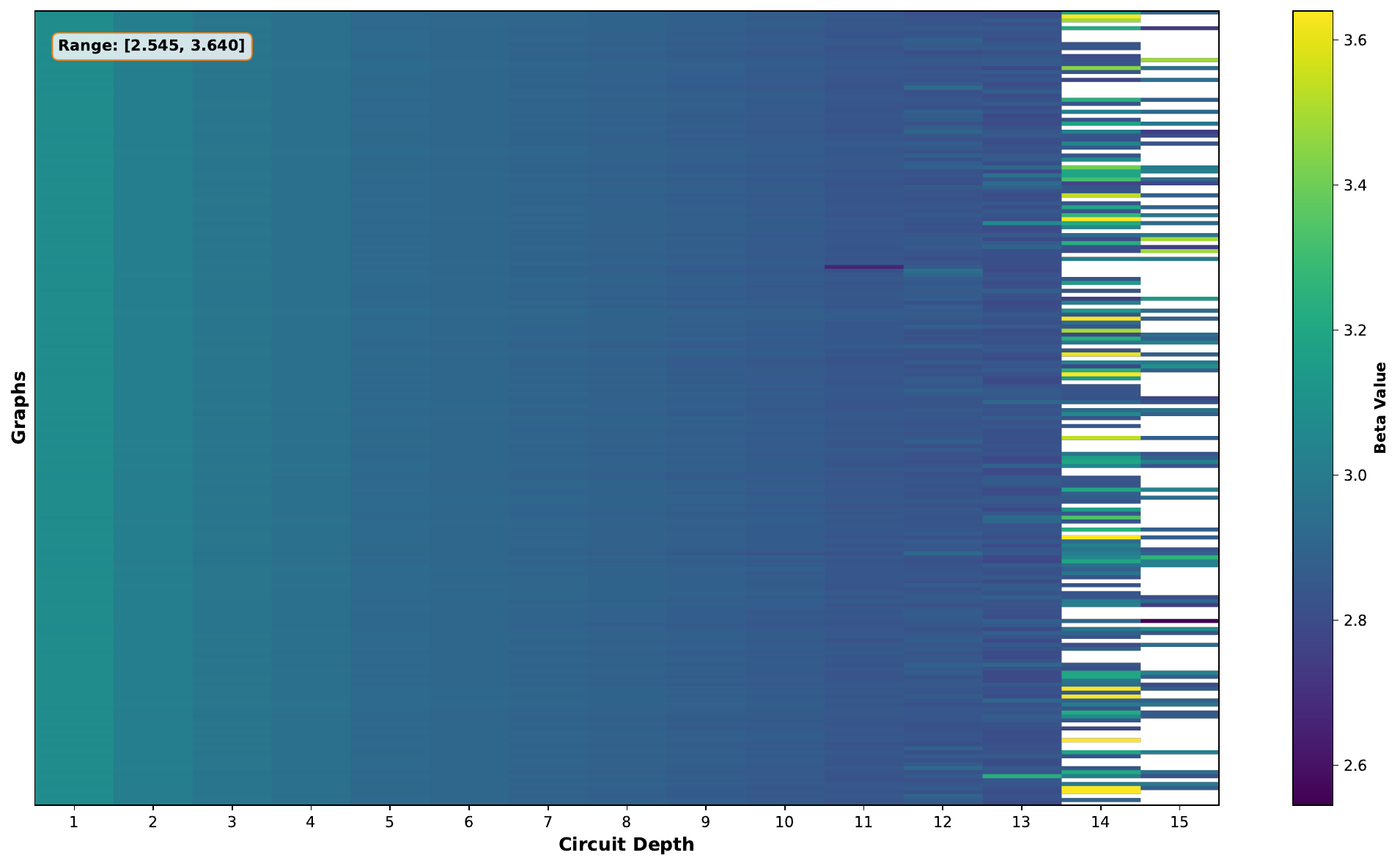}
        \caption{Heatmap of $\beta$ values across instances and depths}
        \label{fig:16n_bheatmap}
    \end{subfigure}
    \caption{Variational parameter distributions for QAOA-GPT generated circuits on 16-qubit spin glass instances at depth \( p_{\text{max}} = 15 \). (a) Mean \( \gamma \) parameters with standard deviation, showing a smooth monotonic decrease across layers. (b) Mean \( \beta \) parameters with standard deviation, exhibiting complex behavior with an initial decrease followed by an upturn at later depths. (c) Heatmap of \( \gamma \) values across all graphs and circuit depths, showing concentration in the range [0.010, 0.623]. (d) Heatmap of \( \beta \) values across all graphs and circuit depths, showing values in the range [2.545, 3.640].}
    \label{fig:16n_parameters}
\end{figure*}

\begin{figure*}[htb]
    \centering
    \begin{subfigure}{0.48\textwidth}
        \centering
        \includegraphics[width=\linewidth]{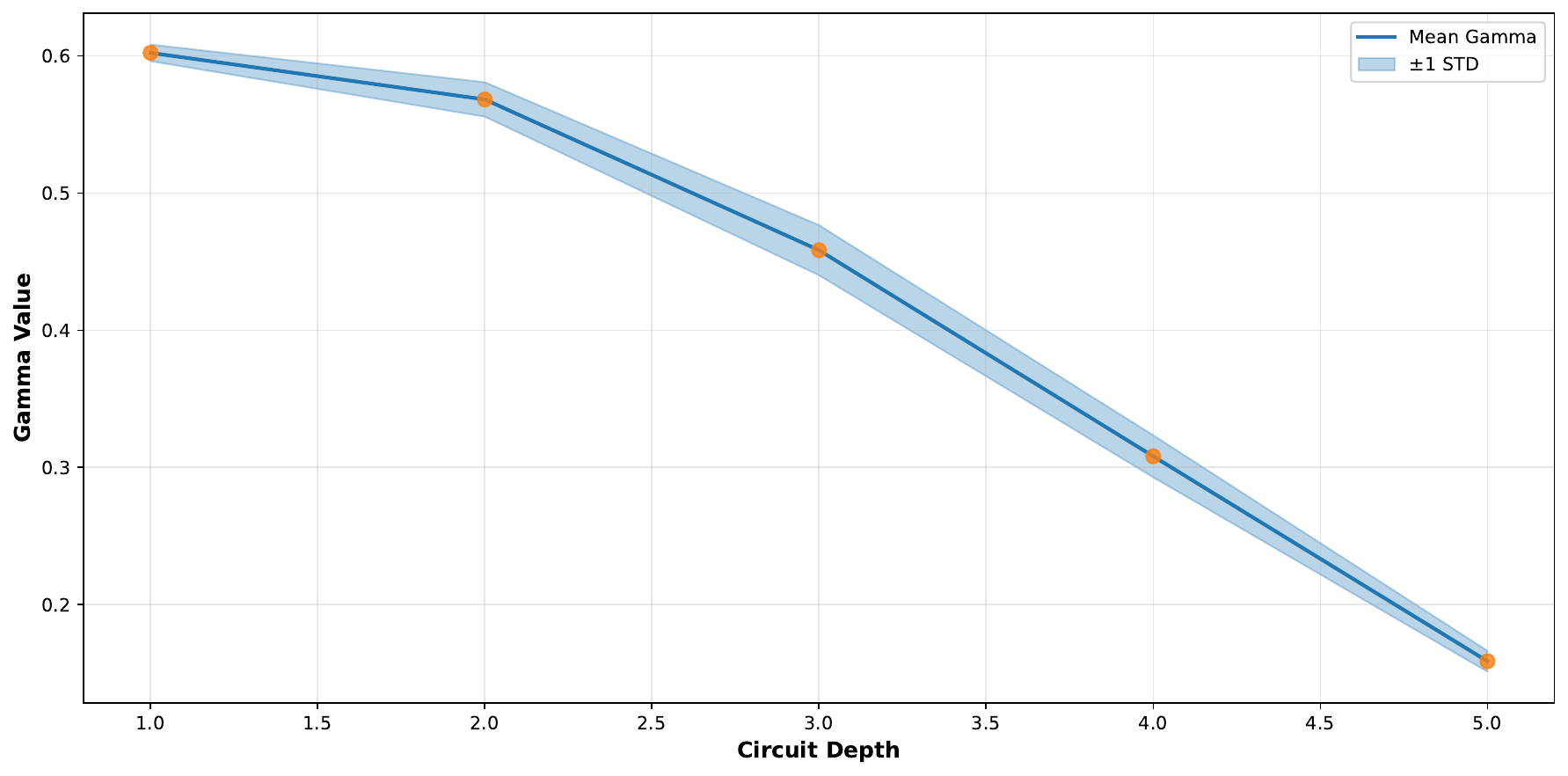}
        \caption{Mean $\gamma$ parameters with standard deviation across circuit depths}
        \label{fig:16n_5l_gscatter}
    \end{subfigure}
    \hfill
    \begin{subfigure}{0.48\textwidth}
        \centering
        \includegraphics[width=\linewidth]{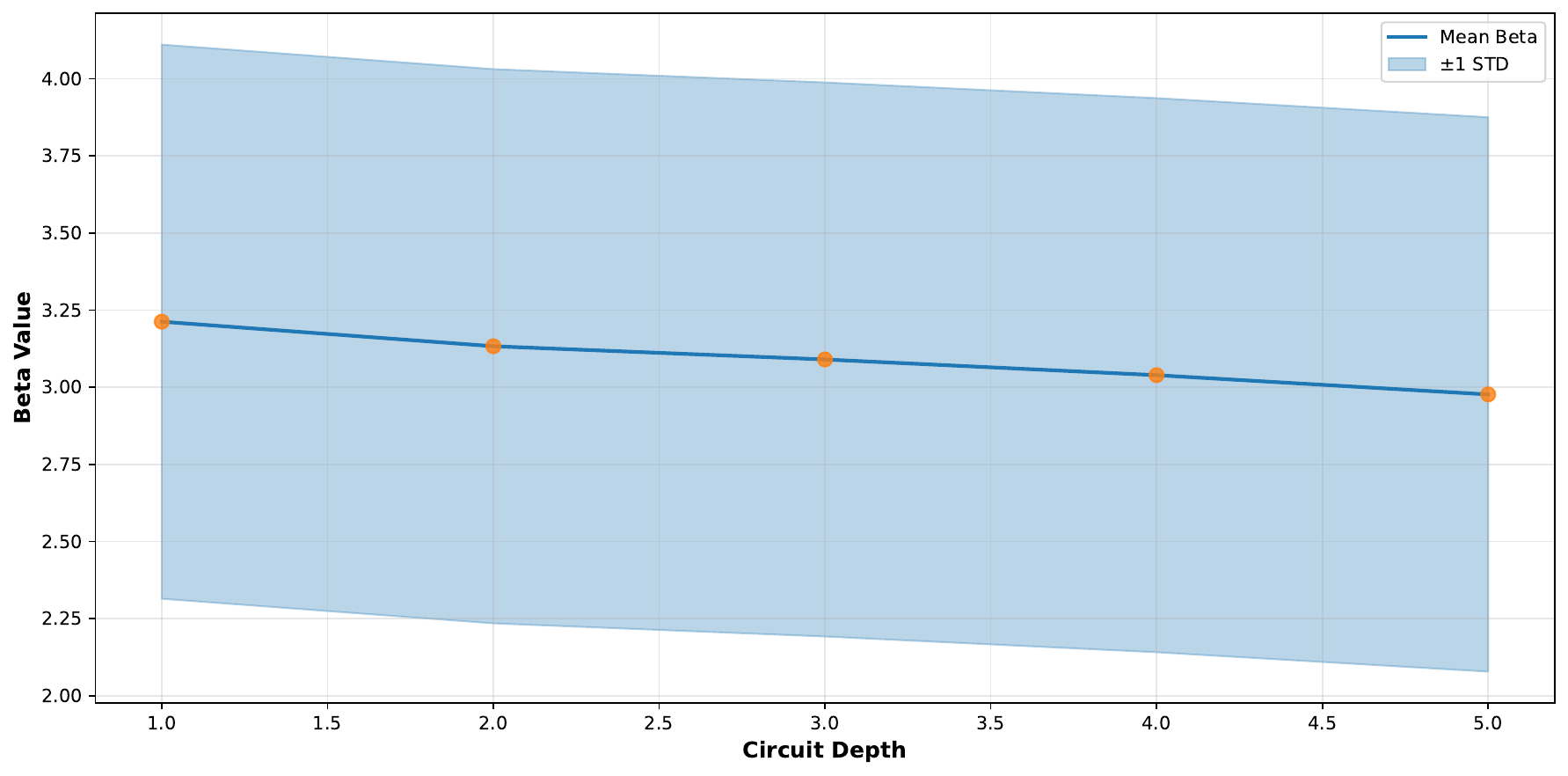}
        \caption{Mean $\beta$ parameters with standard deviation across circuit depths}
        \label{fig:16n_5l_bscatter}
    \end{subfigure}
    \\
    \begin{subfigure}{0.48\textwidth}
        \centering
        \includegraphics[width=\linewidth]{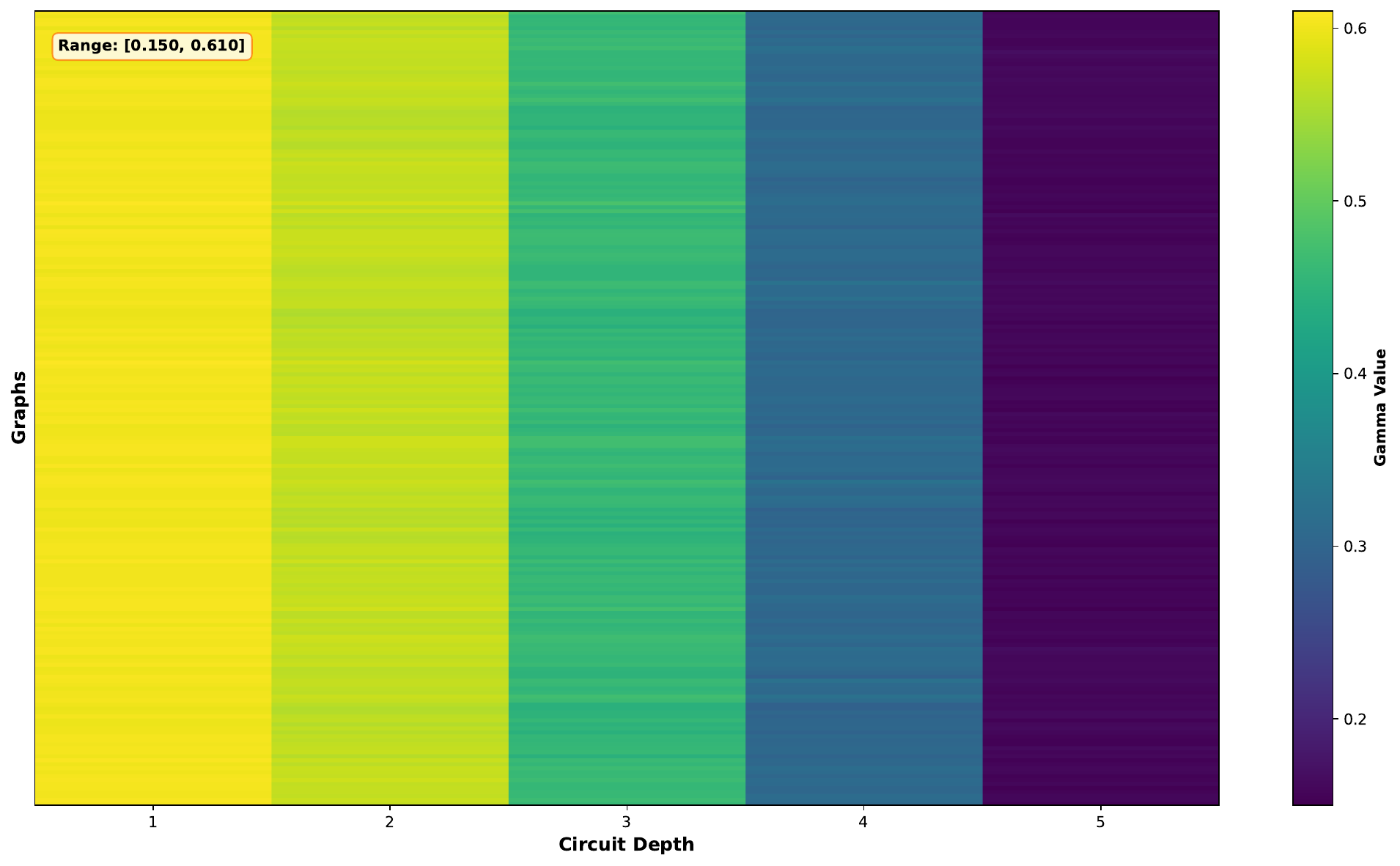}
        \caption{Heatmap of $\gamma$ values across instances and depths}
        \label{fig:16n_5l_gheatmap}
    \end{subfigure}
    \hfill
    \begin{subfigure}{0.48\textwidth}
        \centering
        \includegraphics[width=\linewidth]{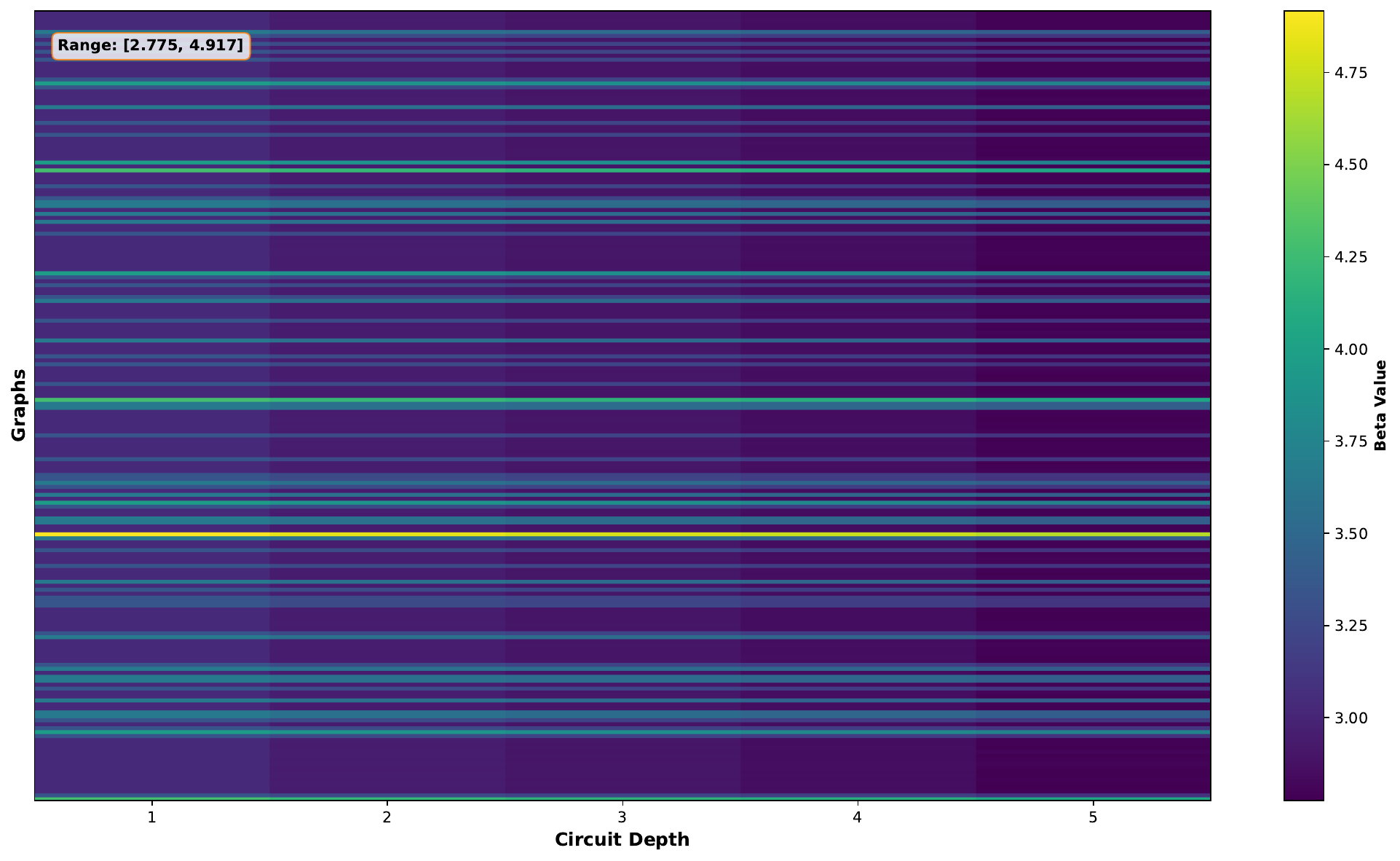}
        \caption{Heatmap of $\beta$ values across instances and depths}
        \label{fig:16n_5l_bheatmap}
    \end{subfigure}
    \caption{Variational parameter distributions for QAOA-GPT generated circuits on 16-qubit spin glass instances at depth \( p_{\text{max}} = 5 \). (a) Mean \( \gamma \) parameters with standard deviation, showing a gentle decrease across layers. (b) Mean \( \beta \) parameters with standard deviation, maintaining a nearly flat profile. (c) Heatmap of \( \gamma \) values across all graphs and circuit depths, showing concentration in the range [0.150, 0.610]. (d) Heatmap of \( \beta \) values across all graphs and circuit depths, showing values in the range [2.775, 4.917].}
    \label{fig:16n_5l_parameters}
\end{figure*}

To gain insight into the model’s internal representation of optimal circuits, we analyze the distributions of the learned $\beta$ and $\gamma$ parameters as functions of circuit depth $p$.
Figures \ref{fig:8n_parameters}, \ref{fig:8n_5l_parameters}, \ref{fig:16n_parameters}, and \ref{fig:16n_5l_parameters} display the average values and one-standard-deviation intervals for each layer, as well as corresponding heatmaps across all test graphs.

The $\beta$ parameters exhibit a smooth oscillatory behavior centered near 3.0 rad, with decreasing variance as circuit depth increases, indicating that the model converges toward stable rotation patterns in deeper layers.
The $\gamma$ parameters, in contrast, cluster near 0.3 rad, consistent with the short-time evolution limit often observed in optimized QAOA circuits.
The narrow spread of both parameters suggests that the model has internalized a consistent structural pattern that generalizes across different problem instances.
Such behavior parallels trends observed in classically optimized ADAPT-QAOA circuits, reinforcing that QAOA-GPT does not merely memorize token sequences but learns physically meaningful parameter relationships.

\subsection{Implications for Circuit Expressivity}

The generative model’s ability to reproduce high-quality circuits without explicit optimization implies an implicit learning of circuit expressivity conditioned on graph topology.
Because the FEATHER embeddings encode global connectivity and higher-order correlations, the transformer effectively learns a compressed representation of the operator pool selection and parameterization process.
This enables the model to generate ans\"atze that respect the structural constraints of ADAPT-QAOA while avoiding irregular circuit topologies that can limit hardware execution efficiency.

The resulting circuits display smooth scaling of depth with system size and consistent layer-wise parameter structure, both of which are desirable for implementation on NISQ hardware.
In practical terms, this indicates that generative circuit synthesis can deliver expressive, hardware-compatible ans\"atze without iterative search, paving the way for automated quantum-algorithm design pipelines.

\subsection{Comparison with Classical Optimization}

While the present study focuses primarily on generative performance, the practical advantage of QAOA-GPT is its computational efficiency.
In conventional ADAPT-QAOA, each circuit layer requires multiple evaluations of gradient operators and repeated classical optimization steps, leading to scaling costs that grow rapidly with both system size and circuit depth.
In contrast, QAOA-GPT generates a complete circuit—including operator sequence and optimized variational parameters—in a single forward pass, reducing computational cost by several orders of magnitude.
Although the generated circuits may not always achieve the absolute best energy found by ADAPT-QAOA, the negligible inference cost makes the generative approach particularly attractive for rapid exploration of large problem spaces or as an initialization scheme for further fine-tuning.

The results presented here demonstrate that QAOA-GPT generalizes effectively to higher-order unconstrained binary optimization problems, achieving near-optimal energies for spin-glass Hamiltonians with cubic terms.
The learned parameter distributions remain stable across circuit depths and system sizes, and the approximation ratios remain consistently above 0.93 for all test cases.
Together, these outcomes validate the hypothesis that generative sequence models can learn the structure of efficient variational quantum circuits directly from data, providing a scalable and data-driven alternative to classical optimization-based variational algorithms.

In the following section, we summarize the implications of these findings for scalable quantum algorithm design and discuss potential extensions of the QAOA-GPT framework to broader classes of optimization and simulation problems.

\section{Conclusions} \label{sec:conclusions}

In this work, we have extended the QAOA-GPT framework of Tyagin \textit{et al.}\ \cite{tyagin2025qaoa} to address HUBO problems, focusing on spin-glass Hamiltonians with cubic interaction terms.
By leveraging FEATHER graph embeddings to represent higher-order connectivity and training on datasets of ADAPT-QAOA circuits, the generative model learns a direct mapping from graph-encoded Hamiltonians to optimized variational circuits.
This approach enables autonomous synthesis of adaptive QAOA-like circuits without the need for iterative classical optimization.

Our results demonstrate that QAOA-GPT maintains high approximation ratios, typically exceeding 0.95 for 16-qubit HUBO instances, while reproducing $\boldsymbol{\beta}, \boldsymbol{\gamma}$ parameter distributions consistent with classically optimized circuits.
The model exhibits robust generalization across system sizes and interaction orders, indicating that it effectively captures the structural correlations between problem topology and circuit design.
Because circuit generation requires only a single forward pass through the network, the computational cost is reduced by several orders of magnitude compared to conventional ADAPT-QAOA workflows, offering a scalable alternative for near-term quantum optimization.

Beyond confirming the adaptability of QAOA-GPT to higher-order cost Hamiltonians, this study underscores the broader promise of generative modeling as a paradigm for quantum algorithm discovery.
Future directions include incorporating reinforcement learning to refine circuit quality through closed-loop feedback, extending the approach to constrained or continuous-variable Hamiltonians, and validating performance on real superconducting and trapped-ion hardware.
More broadly, integrating generative models with physics-informed representations offers a path toward autonomous and data-driven quantum algorithm design, bridging artificial intelligence and quantum optimization in the NISQ era.\\

\begin{acknowledgments}
We acknowledge support by NSF
award DGE-2152168 and DOE award DE-SC0024328. A
portion of the computation for this work was performed on the University of Tennessee Infrastructure for Scientific Applications and Advanced Computing (ISAAC)
computational resources. This
research also used resources of the Oak Ridge Leadership Computing Facility, which is a DOE Office of Science User Facility
supported under Contract DE-AC05-00OR22725.
\end{acknowledgments}

\appendix



\begin{thebibliography}{11}%
	\makeatletter
	\providecommand \@ifxundefined [1]{%
		\@ifx{#1\undefined}
	}%
	\providecommand \@ifnum [1]{%
		\ifnum #1\expandafter \@firstoftwo
		\else \expandafter \@secondoftwo
		\fi
	}%
	\providecommand \@ifx [1]{%
		\ifx #1\expandafter \@firstoftwo
		\else \expandafter \@secondoftwo
		\fi
	}%
	\providecommand \natexlab [1]{#1}%
	\providecommand \enquote  [1]{``#1''}%
	\providecommand \bibnamefont  [1]{#1}%
	\providecommand \bibfnamefont [1]{#1}%
	\providecommand \citenamefont [1]{#1}%
	\providecommand \href@noop [0]{\@secondoftwo}%
	\providecommand \href [0]{\begingroup \@sanitize@url \@href}%
	\providecommand \@href[1]{\@@startlink{#1}\@@href}%
	\providecommand \@@href[1]{\endgroup#1\@@endlink}%
	\providecommand \@sanitize@url [0]{\catcode `\\12\catcode `\$12\catcode
		`\&12\catcode `\#12\catcode `\^12\catcode `\_12\catcode `\%12\relax}%
	\providecommand \@@startlink[1]{}%
	\providecommand \@@endlink[0]{}%
	\providecommand \url  [0]{\begingroup\@sanitize@url \@url }%
	\providecommand \@url [1]{\endgroup\@href {#1}{\urlprefix }}%
	\providecommand \urlprefix  [0]{URL }%
	\providecommand \Eprint [0]{\href }%
	\providecommand \doibase [0]{https://doi.org/}%
	\providecommand \selectlanguage [0]{\@gobble}%
	\providecommand \bibinfo  [0]{\@secondoftwo}%
	\providecommand \bibfield  [0]{\@secondoftwo}%
	\providecommand \translation [1]{[#1]}%
	\providecommand \BibitemOpen [0]{}%
	\providecommand \bibitemStop [0]{}%
	\providecommand \bibitemNoStop [0]{.\EOS\space}%
	\providecommand \EOS [0]{\spacefactor3000\relax}%
	\providecommand \BibitemShut  [1]{\csname bibitem#1\endcsname}%
	\let\auto@bib@innerbib\@empty
	\bibitem [{\citenamefont {Tyagin}\ \emph {et~al.}(2025)\citenamefont {Tyagin},
		\citenamefont {Farag}, \citenamefont {Sherbert}, \citenamefont {Shirali},
		\citenamefont {Alexeev},\ and\ \citenamefont {Safro}}]{tyagin2025qaoa}%
	\BibitemOpen
	\bibfield  {author} {\bibinfo {author} {\bibfnamefont {I.}~\bibnamefont
			{Tyagin}}, \bibinfo {author} {\bibfnamefont {M.~H.}\ \bibnamefont {Farag}},
		\bibinfo {author} {\bibfnamefont {K.}~\bibnamefont {Sherbert}}, \bibinfo
		{author} {\bibfnamefont {K.}~\bibnamefont {Shirali}}, \bibinfo {author}
		{\bibfnamefont {Y.}~\bibnamefont {Alexeev}},\ and\ \bibinfo {author}
		{\bibfnamefont {I.}~\bibnamefont {Safro}},\ }\bibfield  {title} {\bibinfo
		{title} {Qaoa-gpt: Efficient generation of adaptive and regular quantum
			approximate optimization algorithm circuits},\ }\href
	{https://arxiv.org/abs/2504.16350} {\bibfield  {journal} {\bibinfo  {journal}
			{arXiv preprint arXiv:2504.16350}\ } (\bibinfo {year} {2025})}\BibitemShut
	{NoStop}%
	\bibitem [{\citenamefont {Cerezo}\ \emph {et~al.}(2021)\citenamefont {Cerezo},
		\citenamefont {Arrasmith}, \citenamefont {Babbush}, \citenamefont {Benjamin},
		\citenamefont {Endo}, \citenamefont {Fujii}, \citenamefont {McClean},
		\citenamefont {Mitarai}, \citenamefont {Yuan}, \citenamefont {Cincio} \emph
		{et~al.}}]{cerezo2021variational}%
	\BibitemOpen
	\bibfield  {author} {\bibinfo {author} {\bibfnamefont {M.}~\bibnamefont
			{Cerezo}}, \bibinfo {author} {\bibfnamefont {A.}~\bibnamefont {Arrasmith}},
		\bibinfo {author} {\bibfnamefont {R.}~\bibnamefont {Babbush}}, \bibinfo
		{author} {\bibfnamefont {S.~C.}\ \bibnamefont {Benjamin}}, \bibinfo {author}
		{\bibfnamefont {S.}~\bibnamefont {Endo}}, \bibinfo {author} {\bibfnamefont
			{K.}~\bibnamefont {Fujii}}, \bibinfo {author} {\bibfnamefont {J.~R.}\
			\bibnamefont {McClean}}, \bibinfo {author} {\bibfnamefont {K.}~\bibnamefont
			{Mitarai}}, \bibinfo {author} {\bibfnamefont {X.}~\bibnamefont {Yuan}},
		\bibinfo {author} {\bibfnamefont {L.}~\bibnamefont {Cincio}}, \emph
		{et~al.},\ }\bibfield  {title} {\bibinfo {title} {Variational quantum
			algorithms},\ }\href {https://www.nature.com/articles/s42254-021-00348-9}
	{\bibfield  {journal} {\bibinfo  {journal} {Nature Reviews Physics}\ }\textbf
		{\bibinfo {volume} {3}},\ \bibinfo {pages} {625} (\bibinfo {year}
		{2021})}\BibitemShut {NoStop}%
	\bibitem [{\citenamefont {Choquette}\ \emph {et~al.}(2021)\citenamefont
		{Choquette}, \citenamefont {Di~Paolo}, \citenamefont {Barkoutsos},
		\citenamefont {S\'en\'echal}, \citenamefont {Tavernelli},\ and\ \citenamefont
		{Blais}}]{PhysRevResearch.3.023092}%
	\BibitemOpen
	\bibfield  {author} {\bibinfo {author} {\bibfnamefont {A.}~\bibnamefont
			{Choquette}}, \bibinfo {author} {\bibfnamefont {A.}~\bibnamefont {Di~Paolo}},
		\bibinfo {author} {\bibfnamefont {P.~K.}\ \bibnamefont {Barkoutsos}},
		\bibinfo {author} {\bibfnamefont {D.}~\bibnamefont {S\'en\'echal}}, \bibinfo
		{author} {\bibfnamefont {I.}~\bibnamefont {Tavernelli}},\ and\ \bibinfo
		{author} {\bibfnamefont {A.}~\bibnamefont {Blais}},\ }\bibfield  {title}
	{\bibinfo {title} {Quantum-optimal-control-inspired ansatz for variational
			quantum algorithms},\ }\href
	{https://doi.org/10.1103/PhysRevResearch.3.023092} {\bibfield  {journal}
		{\bibinfo  {journal} {Phys. Rev. Res.}\ }\textbf {\bibinfo {volume} {3}},\
		\bibinfo {pages} {023092} (\bibinfo {year} {2021})}\BibitemShut {NoStop}%
	\bibitem [{\citenamefont {Farhi}\ \emph {et~al.}(2014)\citenamefont {Farhi},
		\citenamefont {Goldstone},\ and\ \citenamefont
		{Gutmann}}]{farhi2014quantumapproximateoptimizationalgorithm}%
	\BibitemOpen
	\bibfield  {author} {\bibinfo {author} {\bibfnamefont {E.}~\bibnamefont
			{Farhi}}, \bibinfo {author} {\bibfnamefont {J.}~\bibnamefont {Goldstone}},\
		and\ \bibinfo {author} {\bibfnamefont {S.}~\bibnamefont {Gutmann}},\ }\href
	{https://arxiv.org/abs/1411.4028} {\bibinfo {title} {A quantum approximate
			optimization algorithm}} (\bibinfo {year} {2014}),\ \Eprint
	{https://arxiv.org/abs/1411.4028} {arXiv:1411.4028 [quant-ph]} \BibitemShut
	{NoStop}%
	\bibitem [{\citenamefont {Blekos}\ \emph {et~al.}(2024)\citenamefont {Blekos},
		\citenamefont {Brand}, \citenamefont {Ceschini}, \citenamefont {Chou},
		\citenamefont {Li}, \citenamefont {Pandya},\ and\ \citenamefont
		{Summer}}]{blekos2024review}%
	\BibitemOpen
	\bibfield  {author} {\bibinfo {author} {\bibfnamefont {K.}~\bibnamefont
			{Blekos}}, \bibinfo {author} {\bibfnamefont {D.}~\bibnamefont {Brand}},
		\bibinfo {author} {\bibfnamefont {A.}~\bibnamefont {Ceschini}}, \bibinfo
		{author} {\bibfnamefont {C.-H.}\ \bibnamefont {Chou}}, \bibinfo {author}
		{\bibfnamefont {R.-H.}\ \bibnamefont {Li}}, \bibinfo {author} {\bibfnamefont
			{K.}~\bibnamefont {Pandya}},\ and\ \bibinfo {author} {\bibfnamefont
			{A.}~\bibnamefont {Summer}},\ }\bibfield  {title} {\bibinfo {title} {A review
			on quantum approximate optimization algorithm and its variants},\ }\href
	{https://doi.org/10.1016/j.physrep.2024.03.002} {\bibfield  {journal}
		{\bibinfo  {journal} {Physics Reports}\ }\textbf {\bibinfo {volume} {1068}},\
		\bibinfo {pages} {1} (\bibinfo {year} {2024})}\BibitemShut {NoStop}%
	\bibitem [{\citenamefont {Zhu}\ \emph {et~al.}(2022)\citenamefont {Zhu} \emph
		{et~al.}}]{zhu2022adaptqaoa}%
	\BibitemOpen
	\bibfield  {author} {\bibinfo {author} {\bibfnamefont {L.}~\bibnamefont
			{Zhu}} \emph {et~al.},\ }\bibfield  {title} {\bibinfo {title} {Adaptive
			quantum approximate optimization algorithm for solving combinatorial problems
			on a quantum computer},\ }\href
	{https://journals.aps.org/prresearch/abstract/10.1103/PhysRevResearch.4.033029}
	{\bibfield  {journal} {\bibinfo  {journal} {Physical Review Research}\
		}\textbf {\bibinfo {volume} {4}},\ \bibinfo {pages} {043178} (\bibinfo {year}
		{2022})}\BibitemShut {NoStop}%
	\bibitem [{\citenamefont {Jain}\ \emph {et~al.}(2022)\citenamefont {Jain},
		\citenamefont {Coyle}, \citenamefont {Kashefi},\ and\ \citenamefont
		{Kumar}}]{jain2022graph}%
	\BibitemOpen
	\bibfield  {author} {\bibinfo {author} {\bibfnamefont {N.}~\bibnamefont
			{Jain}}, \bibinfo {author} {\bibfnamefont {B.}~\bibnamefont {Coyle}},
		\bibinfo {author} {\bibfnamefont {E.}~\bibnamefont {Kashefi}},\ and\ \bibinfo
		{author} {\bibfnamefont {N.}~\bibnamefont {Kumar}},\ }\bibfield  {title}
	{\bibinfo {title} {Graph neural network initialisation of quantum approximate
			optimisation},\ }\href {https://quantum-journal.org/papers/q-2022-11-17-861/}
	{\bibfield  {journal} {\bibinfo  {journal} {Quantum}\ }\textbf {\bibinfo
			{volume} {6}},\ \bibinfo {pages} {861} (\bibinfo {year} {2022})}\BibitemShut
	{NoStop}%
	\bibitem [{\citenamefont {Rozemberczki}\ and\ \citenamefont
		{Sarkar}(2020)}]{rozemberczki2020characteristic}%
	\BibitemOpen
	\bibfield  {author} {\bibinfo {author} {\bibfnamefont {B.}~\bibnamefont
			{Rozemberczki}}\ and\ \bibinfo {author} {\bibfnamefont {R.}~\bibnamefont
			{Sarkar}},\ }\bibfield  {title} {\bibinfo {title} {Characteristic functions
			on graphs: Birds of a feather, from statistical descriptors to parametric
			models},\ }in\ \href {https://dl.acm.org/doi/abs/10.1145/3340531.3411866}
	{\emph {\bibinfo {booktitle} {Proceedings of the 29th ACM international
				conference on information \& knowledge management}}}\ (\bibinfo {year}
	{2020})\ pp.\ \bibinfo {pages} {1325--1334}\BibitemShut {NoStop}%
	\bibitem [{\citenamefont {Karpathy}()}]{Karpathy}%
	\BibitemOpen
	\bibfield  {author} {\bibinfo {author} {\bibnamefont {Karpathy}},\ }\href
	{https://github.com/karpathy/nanoGPT.git} {\bibinfo {title} {nano{G}{P}{T}:
			The simplest, fastest repository for training/finetuning medium-sized
			gpts.}}\BibitemShut {Stop}%
	\bibitem [{\citenamefont {Vaswani}\ \emph {et~al.}(2017)\citenamefont
		{Vaswani}, \citenamefont {Shazeer}, \citenamefont {Parmar}, \citenamefont
		{Uszkoreit}, \citenamefont {Jones}, \citenamefont {Gomez}, \citenamefont
		{Kaiser},\ and\ \citenamefont {Polosukhin}}]{NIPS2017_3f5ee243}%
	\BibitemOpen
	\bibfield  {author} {\bibinfo {author} {\bibfnamefont {A.}~\bibnamefont
			{Vaswani}}, \bibinfo {author} {\bibfnamefont {N.}~\bibnamefont {Shazeer}},
		\bibinfo {author} {\bibfnamefont {N.}~\bibnamefont {Parmar}}, \bibinfo
		{author} {\bibfnamefont {J.}~\bibnamefont {Uszkoreit}}, \bibinfo {author}
		{\bibfnamefont {L.}~\bibnamefont {Jones}}, \bibinfo {author} {\bibfnamefont
			{A.~N.}\ \bibnamefont {Gomez}}, \bibinfo {author} {\bibfnamefont {L.~u.}\
			\bibnamefont {Kaiser}},\ and\ \bibinfo {author} {\bibfnamefont
			{I.}~\bibnamefont {Polosukhin}},\ }\bibfield  {title} {\bibinfo {title}
		{Attention is all you need},\ }in\ \href
	{https://proceedings.neurips.cc/paper_files/paper/2017/file/3f5ee243547dee91fbd053c1c4a845aa-Paper.pdf}
	{\emph {\bibinfo {booktitle} {Advances in Neural Information Processing
				Systems}}},\ Vol.~\bibinfo {volume} {30},\ \bibinfo {editor} {edited by\
		\bibinfo {editor} {\bibfnamefont {I.}~\bibnamefont {Guyon}}, \bibinfo
		{editor} {\bibfnamefont {U.~V.}\ \bibnamefont {Luxburg}}, \bibinfo {editor}
		{\bibfnamefont {S.}~\bibnamefont {Bengio}}, \bibinfo {editor} {\bibfnamefont
			{H.}~\bibnamefont {Wallach}}, \bibinfo {editor} {\bibfnamefont
			{R.}~\bibnamefont {Fergus}}, \bibinfo {editor} {\bibfnamefont
			{S.}~\bibnamefont {Vishwanathan}},\ and\ \bibinfo {editor} {\bibfnamefont
			{R.}~\bibnamefont {Garnett}}}\ (\bibinfo  {publisher} {Curran Associates,
		Inc.},\ \bibinfo {year} {2017})\BibitemShut {NoStop}%
	\bibitem [{\citenamefont {{NVIDIA Corporation}}()}]{nvidia_cudaq}%
	\BibitemOpen
	\bibfield  {author} {\bibinfo {author} {\bibnamefont {{NVIDIA
					Corporation}}},\ }\href@noop {} {\bibinfo {title} {{NVIDIA CUDA-Q}
			framework}},\ \bibinfo {howpublished}
	{\url{https://github.com/NVIDIA/cuda-quantum}}\BibitemShut {NoStop}%
\end{thebibliography}

%

\end{document}